# Floor Raiser or Ceiling Limiter? Differential Storytelling Outcomes with a Child-Centric GenAI System Across Individual Differences

**1. Min Fan**[1,2,*]
[1]School of Animation and Digital Arts, Communication University of China, Beijing, China
[2] Media Laboratory, Massachusetts Institute of Technology, Cambridge, MA, USA
ORCID: 0000-0001-5323-8168
Email: mfan@cuc.edu.cn
**2. Wanqing Ma**
School of Animation and Digital Arts, Communication University of China, Beijing, China
ORCID: 0009-0000-0262-1562
Email: mawanqing@cuc.edu.cn
**3. Xinyue Cui**[1,2]
[1]School of Animation and Digital Arts, Communication University of China, Beijing, China
[2] Graduate School of Education, University of Pennsylvania, Philadelphia, PA, USA
ORCID: 0009-0000-3664-7138
Email: cuixinyue152@gmail.com
**4. Xiaolu Dai**
Beijing Normal-Hong Kong Baptist University, Faculty of Arts and Social Sciences, Zhuhai, Guangdong, China
ORCID: 0000-0002-0442-8130
Email: xxd53@connect.hku.hk
**5. Shengyu Huang**
Faculty of Arts and Sciences, Harvard University, Cambridge, MA, USA
ORCID: 0009-0006-4376-4977
Email: shengyuhuang@fas.harvard.edu

*Corresponding Author: **Min Fan** (mfan@cuc.edu.cn, mfan1028@media.mit.edu)

# Abstract

Generative AI (GenAI) holds promise for democratizing creative literacy, yet whether it benefits all children equally remains unclear. Using a child-centric GenAI storytelling system for children aged 7–12, we conducted a mixed-methods within-subjects experiment (N = 40, Grades 2–6) comparing GenAI-assisted and traditional storyboard conditions. Three findings emerged. First, the GenAI-assisted condition was associated with a floor-raising convergence pattern, with the quality gap narrowing by 83.5%, driven by lower-end support and upper-end constraint mechanisms. This convergence was dimension-selective, improving creativity and richness while leaving coherence and narrative structure tied to baseline performance. Second, younger children more often selected semantically distant keywords while older children preferred semantically closer ones, although engagement orientation varied across individuals regardless of age. Third, image regeneration was positively associated with structural quality dimensions, though this association was attenuated after baseline control. We propose mechanism-contingent scaffolding as a design principle for adaptive GenAI storytelling systems serving diverse children.

# 1. Introduction

Creative storytelling is the process where children employ narrative structure, mental imagery, and oral language to construct and communicate imaginative worlds (Peck, 1989). It plays a central role in elementary literacy education, fostering language development (Garzotto et al., 2010; Peck, 1989), creativity (Zarei et al., 2020), social-emotional growth (Koivula et al., 2020), and narrative thinking (Garcia & Rossiter, 2010). In school settings, creative storytelling serves different instructional goals across grades. Younger learners are primarily developing basic literacy and learning to retell simple personal experiences, while older elementary students are expected to construct more complex, multi-scene narratives with developed characters, emotional arcs, and coherent plot (Applebee, 1978; Fan et al., 2024; L. Zhang, 2019). Any tool intended to support creative storytelling in schools must therefore be designed to accommodate this wide developmental range and educational requirements. Traditional approaches to classroom storytelling involve paper-based storyboards, drawing activities, and guided dialogues, which provide valuable scaffolding through audio-visual and kinesthetic engagement (Catala et al., 2017; Kleeck et al., 2003; Z. Zhang et al., 2022). However, these tools typically offer limited adaptive and personalized support and rely heavily on experienced instructors to provide real-time, individualized feedback to each student.

The emergence of generative AI (GenAI)[1] offers new possibilities for supplementing these traditional approaches, supporting more dynamic, personalized, and effective storytelling. Recent research in Educational Technology (EduTech) and Human-Computer-Interaction (HCI) communities has explored *child-centric GenAI storytelling tools*[2], including visual storytelling (Bernal et al., 2019; Han & Cai, 2023), parent–child narrative dialogue (Sun et al., 2024; Z. Zhang et al., 2022), and multi-modal creative storytelling assistance (Ye et al., 2025; C. Zhang et al., 2021). These systems have demonstrated that AI-generated content such as images, sketches, and textual prompts can enhance children's creative engagement and story quality (Sun et al., 2024; Ye et al., 2025; C. Zhang et al., 2021, 2024). However, existing research has largely evaluated AI storytelling tools at the aggregate level, without systematically examining how children's different individual characteristics (e.g., varying creative storytelling baselines, developmental stages) moderate the degree of AI benefit, or whether AI's effects operate uniformly across different dimensions of narrative quality. Although a few studies have noted age-related variation in children's AI interactions, such as differences in response elaboration across age groups (Liu et al., 2025) and developmental thresholds for narrative transfer after AI interaction (Ye et al., 2025), these observations concern interaction behaviors rather than differential quality outcomes. In a typical elementary school context, children's narrative capabilities span a wide range, from those who struggle to compose a simple sequence of events in lower grades to those who independently construct multi-layered narratives with complex character psychology in higher grades (Fan et al., 2024; L. Zhang, 2019). Understanding how GenAI creative storytelling tools affect children across this developmental spectrum is essential for informed educational deployment, whether through adaptive system design that adjusts scaffolding by developmental stage, differentiated classroom integration guided by teachers, or a combination of both.

While direct evidence from children is lacking, a growing body of research with adult populations offers a useful reference point (Brynjolfsson et al., 2025; Cen & Shakibaei, 2026; Dell'Acqua et al., 2023; Doshi & Hauser, 2024). In creative writing, lower-ability adult writers tend to benefit disproportionately from AI-generated ideas, while the effects on higher-ability writers remain contested, ranging from negligible benefit to potential quality decline depending on the level of guidance provided (Cen & Shakibaei, 2026; Doshi & Hauser, 2024). This pattern mirrors broader findings from other AI-assisted professional contexts, where lower-performing individuals tend to gain the most from GenAI support (Brynjolfsson et al., 2025; Dell'Acqua et al., 2023), though the direction and magnitude of AI benefit can depend on task characteristics and users' ability to critically evaluate AI output (Dell'Acqua et al.,

[1] In this paper, *GenAI* refers to large language models and multimodal generation models capable of producing text, images, or other media from user inputs.

[2] By *child-centric GenAI storytelling tools*, we refer to systems that integrate GenAI capabilities into interfaces and interactions purposefully designed to accommodate children's developmental needs and educational contexts, distinguishing them from general-purpose GenAI tools not specifically designed for children's use.

2023). Whether similar equalizing patterns hold for children, whose developmental constraints differ fundamentally from adult skill gaps, and through what process mechanisms they arise, remains an open question that this study aims to address.

In this study, we investigate how child-centric GenAI storytelling assistance affects children across individual differences. We operationalize "individual differences" through three dimensions: (1) *baseline storytelling performance*, measured as within-age relative performance by expert raters against grade-level expectations; (2) *developmental stage*, operationalized as grade level; and (3) *self-assessed creative writing ability* as a secondary confidence measure. Gender is reported descriptively but not analyzed inferentially due to sample imbalance. We draw on data from a mixed-methods within-subjects experiment with 40 elementary children (38 with complete paired data; Grades 2–6) using *StoryPrompt*, an AI-empowered storytelling system that we previously developed and evaluated (Fan et al., 2024, 2025). In this system, children first plan a story structure (selecting characters, scenes, and emotional arcs), then compose six story paragraphs through storytelling, each anchored by an AI-generated keyword chosen from a set of strongly and weakly associated options generated in real time based on children's storytelling performance. Once each paragraph is complete, the system produces a comic-style AI-generated image to capture the story content.

Our prior evaluation established that the AI-assisted storytelling significantly improved overall story quality compared to traditional storyboarding, with gains concentrated in creativity and richness (Fan et al., 2025). This study goes beyond these overall effects. Through expanded quantitative analysis and substantially deepened qualitative coding of observational data, interviews, and storytelling artifacts for the participants, we further examine for whom AI assistance is most beneficial, in which quality dimensions its effects operate, and through what process mechanisms children with different characteristics engage with AI features differently. We address three research questions (RQs):

- **RQ1** (Outcomes): How do storytelling quality outcomes differ between the AI-assisted and traditional storyboard conditions for children with varying baseline performance levels, and in which quality dimensions?
- **RQ2** (Process): What engagement mechanisms do children with different characteristics exhibit when interacting with AI-generated keywords and images during the storytelling process?
- **RQ3** (Process-outcome link): How do children's engagement strategies with GenAI features relate to their storytelling quality outcomes?

This study yields three main findings. First, the GenAI-assisted condition was associated with a floor-raising convergence pattern. Lower-performing children gained substantially in creativity and richness through two lower-end support mechanisms, while higher-performing children experienced selective decreases through two upper-end constraint patterns, substantially reducing the quality gap. This equalization was dimension-selective, improving content dimensions while leaving structural dimensions largely dependent on baseline performance. Second, exploratory quantitative trends and qualitative cases suggested that younger children more often selected weakly associated (semantically distant) keywords while older children preferred strongly associated ones, although neither keyword type independently predicted quality. Children's engagement orientation, their openness to incorporating novel information versus adhering to a preconceived plan, varied across individuals regardless of age. Third, image regeneration was positively associated with coherence and narrative structure, though this association was largely attenuated when baseline performance was controlled, indicating a tentative link that warrants larger-sample validation. We note that our analyses focus on immediate storytelling quality and observable engagement strategies; whether these patterns reflect long-term skill development would require longitudinal investigation.

This study contributes: (1) early empirical evidence of floor-raising convergence in children's AI-assisted creative storytelling, extending adult findings to an ecologically valid elementary school context; (2) evidence that this convergence is dimension-selective, with process mechanisms linking engagement behaviors to quality outcomes; and (3) design considerations for mechanism-contingent scaffolding in adaptive GenAI storytelling systems serving diverse children.

## 2. Related Work

We review child-centric AI storytelling systems, then examine AI's differential effects across varying ability levels in both adult and children's contexts, and summarize approaches to evaluating creative storytelling quality.

### 2.1 Child-Centric AI Storytelling Systems and Children's Behavioral Engagement

Research in HCI and educational technology has produced a growing body of child-centric AI storytelling systems, progressing from traditional AI techniques to generative AI capabilities. These systems can be broadly characterized by the type of creative scaffolding provided: *visual content generation*, *dialogic interaction*, or *a combination of both*. While AI-generated stimuli enhance children's creative engagement at the aggregate level, important variations in engagement patterns and differential effects remain unresolved.

*Visual and multimodal generation systems* have explored how AI-produced imagery can scaffold children's storytelling. Bernal et al. (2019) designed *Paper Dreams*, converting children's text into real-time AI images to encourage divergent thinking and narrative ownership, though its single-page format limits support for sequential storytelling. C. Zhang et al. (2022) extended this approach with *StoryDrawer*, which uses sketch recognition and neural sketch generation to transform children's oral narratives into drawings. Their evaluation revealed two distinct behavioral patterns: *narrative-first storytelling* (developing stories through narration while using AI drawing support) and *drawing-first storytelling* (enriching visual elements by actively triggering AI sketch generation), with drawing-first children producing more scenes but shorter descriptions per event. However, the study did not link these patterns to children's age, baseline storytelling performance, or overall story quality. Han and Cai (2023) addressed a key tension in AI-generated visual content through *AIStory*, which pairs Stable Diffusion with a constrained sticker-based interface to balance generative power with children's creative agency. While AI generates rich visual content, unconstrained generation risks reducing children's independent creative thinking.

*LLM-powered dialogue and scaffolding systems* have leveraged large language models to provide richer, more context-sensitive textual support. Z. Zhang et al. (2022) designed *StoryBuddy*, which uses GPT-3 to generate interactive question-and-answer sessions during parent-child storytelling. AI-generated questions reduced parents' burden and stimulated children's divergent associations, though the system's effectiveness depends on parental mediation. C. Zhang et al. (2024) presented *Mathemyths*, an LLM-powered agent embedding mathematical language into co-creative storytelling. Whether such domain-specific scaffolding generalizes to open-ended creative storytelling remains untested. More recently, Ye et al. (2025) developed *Colin*, a multimodal system that uses GPT-4 and ChatGLM for question-scaffolded story co-creation. Their evaluation noted an age-related pattern. Only the youngest participants (aged 5–6) were unable to construct new stories after AI interaction, while older children successfully transferred narrative concepts. This finding is suggestive of developmental differences in how children leverage AI scaffolding, yet age was not analyzed as a primary variable. Sun et al. (2024) presented *StoryChat*, simulating story characters via LLMs for children's dialogic reading. The AI's role shifts from content generator to conversational partner. Kim et al. (2024) designed *TinyTeller*, generating personalized adaptive stories for children's profiles, though its effects on storytelling quality remain unevaluated.

*Comparative studies* between AI and human or non-AI storytelling conditions provide further insight into both the benefits and boundaries of current AI scaffolding. Chin et al. (2024) compared children aged 5–6 co-creating stories with ChatGPT versus a parent, finding more conversational turns in parent-child interactions with parents asking more about characters and settings. Although children preferred ChatGPT, the AI's structured responses limited back-and-forth exchanges. Liu et al. (2025) extended this line of inquiry with a larger sample of 73 children aged 4–8, finding AI-condition children produced more brief responses and fewer detailed ones. Approximately 51% of detailed AI responses consisted of sequential action descriptions alone, whereas human-partnered children enriched responses with contextualization and causal reasoning. Neither study assessed whether children's baseline performance modulated these patterns or whether engagement differences affected story quality. In prior work, we also designed and evaluated *StoryPrompt*, an AI-powered storytelling system, against a traditional storyboard method with 40 children from Grades 2–6 and found that AI co-creation significantly improved storytelling creativity and richness at the aggregate level, with children demonstrating more purposeful planning and strategic use of AI-generated content (Fan et al., 2025). Yet this evaluation also did not disaggregate outcomes by children's baseline performance or developmental stage, leaving open the question of whether the observed benefits were uniformly distributed. Therefore, existing evaluations share a critical limitation in analyzing outcomes at the aggregate level. Whether benefits are shared equally or concentrated among children with particular characteristics remains the question this study addresses.

## 2.2 AI's Differential Effects on Users with Varying Ability Levels

Children's narrative capabilities span a wide developmental range shaped by factors that could moderate how AI assistance benefits individual learners. Grade level and age consistently predict narrative quality, with older children producing structurally complex stories while younger children develop basic literacy and retelling ability (Applebee, 1978; Fan et al., 2024; L. Zhang, 2019). Beyond developmental stage, children's writing self-efficacy (i.e., the confidence in their own writing ability) has been shown to predict text quality independently of objective skill, with higher self-efficacy associated with greater motivation and creative risk-taking (Cordeiro et al., 2018). Gender differences are documented, with girls tending to score higher in elementary school, though effect sizes vary (Cordeiro et al., 2018; Reilly et al., 2019). As reviewed in Section 2.1, several child-AI storytelling evaluations have noted variation along some of these dimensions, such as age-related differences in response elaboration (Liu et al., 2025) and developmental thresholds for narrative concept transfer (Ye et al., 2025), but these observations concern interaction behaviors rather than differential quality outcomes, and whether AI's effects on quality vary across individual characteristics remains unexamined.

Research with adult populations provides a reference point. Doshi and Hauser (2024) conducted a large-scale experiment in which approximately 300 adults wrote short stories with or without access to GPT-4-generated ideas. AI-assisted stories were rated as more creative and enjoyable, but the improvement was concentrated among writers with the lowest baseline creativity scores who saw gains of over 10% in novelty while higher-creativity writers showed minimal additional benefit. AI-assisted stories were also more similar to each other, suggesting a trade-off between individual enhancement and collective diversity. Cen and Shakibaei (2026) examined a related dimension by comparing high-AI-support, low-AI-support, and control conditions in creative writing tasks. Both AI groups outperformed the control, with high-AI support producing the largest gains in creativity by alleviating the cognitive burden of technical writing aspects, while low-AI support fostered greater learner independence through the interplay between AI and instructor feedback. This distinction in scaffolding levels is relevant to child-centric tool design, where the degree and type of built-in support may shape the distribution of benefits across children with varying baselines.

The *equalizing* pattern where lower-performing individuals benefit disproportionately from AI has also been documented in professional contexts beyond creative writing (Doshi & Hauser, 2024). Brynjolfsson et al. (2025) found that AI-assisted customer service agents experienced a 34% productivity improvement among the lowest-performing workers with minimal impact on highest-skilled workers while Dell'Acqua et al. (2023) reported similar findings among management consultants using GPT-4. However, Dell'Acqua et al. also found that for tasks outside AI's capability frontier, AI use decreased performance, suggesting the equalizing effect depends on task nature.

These studies share a common framework: AI benefits lower-performing individuals by providing resources they lack. An alternative mechanism, however, has received less attention. Cognitive Load Theory (Sweller, 1988) suggests that when working memory is divided across concurrent demands, performance on each task suffers. In multimodal creative tasks such as illustrated storytelling, children must simultaneously manage visual production and narrative composition, creating potential resource competition between modalities. Recent work in visual art education has suggested that AI image generation tools may reduce the procedural demands of technical execution, potentially reallocating cognitive resources toward creative exploration (Bian et al., 2025), though this reallocation has not been empirically isolated from other benefits of AI assistance. Whether a similar resource reallocation mechanism operates in children's multimodal storytelling, where AI-generated images might liberate cognitive resources otherwise consumed by hand-drawing, has not been investigated. Transferring adult findings to children is not straightforward, as developmental factors produce more pronounced variation. Whether this equalizing pattern generalizes to children's AI-assisted storytelling is a central question of this study.

## 2.3 Evaluating Creative Storytelling Quality

Creativity has multiple theoretical lenses. Boden (1990) distinguished between historical creativity (ideas never before conceived by anyone) and psychological creativity (ideas new to the individual), particularly relevant for children, where novelty is assessed relative to the child's developmental context. Rhodes' (1961) 4Ps framework (i.e., Product, Process, Person, and Press) highlights attention to the story produced, the cognitive strategies used, the creator characteristics, and the available tools. Kupers et al. (2019) in a systematic review of children's creativity research, argued for shifting from macro-static assessments toward micro-level process measures capturing children's dynamic creative strategies in technological environments, directly relevant to AI-assisted storytelling where creativity unfolds through interactions with features like keyword selection and image generation.

In children's creative storytelling research, evaluations span product and process levels. At the *product* level, the Consensual Assessment Technique (CAT) developed by Amabile (1982) is widely regarded as the most appropriate approach for evaluating creative products (Baer & McKool, 2009; Hennessey & Amabile, 2010). CAT uses independent expert ratings as the operational definition of creativity. Building on CAT, Chu et al. (2015) employed five expert raters to assess children's storytelling quality across four dimensions including *creativity*, *fluency*[3], *richness*, and *story structure* in the context of an enactment-based animated storytelling system, providing a precedent for multi-dimensional assessment in child-technology interaction research. Elgarf et al. (2022) adopted a related approach in evaluating children's robot-assisted storytelling across *fluency*, *flexibility*, *elaboration*, and *originality*. Zarei et al. (2020) investigated how self-avatars and story-relevant avatars affect children's storytelling, connecting quality ratings with behavioral observations. At the *process* level, several of the AI storytelling evaluations have captured behavioral indicators of creative engagement, such as narrative-first versus drawing-first patterns (C. Zhang et al., 2022) and analysis of response styles during story co-creation (Liu et al., 2025), though these have not been linked to quality outcomes or individual differences.

The adult evidence suggests that AI's effects are not uniform across creative output, with content-oriented dimensions (e.g., creativity, richness) potentially responding differently than structure-oriented ones (e.g., coherence, story structure). We adopt a multi-dimensional CAT approach with blind expert ratings. Following Rhodes' (1961) 4Ps framework, we analyze expert-rated story quality (Product), children's engagement with AI features (Process), baseline performance and developmental stage (Person), and AI-versus-traditional comparison (Press). This integration enables examination of whether AI's effects are dimension-selective, interact with individual differences, and how children's AI engagement strategies relate to quality outcomes.

# 3. Methods

We conducted a study with 40 children (aged 7-12) in a local elementary school, who used both the traditional storyboard method (Storyboard condition) and a GenAI system we designed (AI condition) to complete their creative storytelling tasks. We employed a counterbalanced within-subjects design for two reasons: first, it controls for individual differences including age, grade, gender, writing ability, and prior experience that could otherwise confound between-condition comparisons, enabling cleaner inferences about the effect of AI assistance on story quality; second, it ensures that all participants experience the AI condition, promoting educational equity.

[3] In the creativity research tradition, *fluency* typically refers to the quantity and speed of idea generation, i.e., how many different ideas an individual can produce in a given time (Guilford, 1967; Torrance, 1974). In the context of storytelling assessment, Chu et al. (2015) adapted this term to evaluate narrative flow and readability rather than ideational quantity.

## 3.1 Research Instrument: A GenAI-Empowered Creative Storytelling System

We first designed an AI-empowered creative storytelling system called *StoryPrompt* for elementary school children aged 7–12, implemented in the Unity game engine. Children co-create illustrated storybooks through two phases: story planning and story creation (Fig. 1).

In the story planning phase, the child first selects a main character from a curated library of ten characters spanning four types, *talking animals or objects* (e.g., a cat, a robot*), characters with magical or supernatural abilities* (e.g., a witch, a wizard), *science-fiction characters* (e.g., a scientist, an astronaut), and *ordinary children*. The child then assigns three personality traits to the character and selects three sequential story scenes from a library of seven settings across four categories: *natural environments* (e.g., forest, volcano), *fantasy locations* (e.g., castle, cabin), *science-fiction settings* (e.g., laboratory, outer space), and *everyday school environments*. All these types were designed to reflect the genre preferences identified in our formative study (Fan et al., 2024). Finally, the child assigns one emotion to each scene, selecting from the six basic options involving *positive*, *negative*, and *neutral* ones. This planning phase establishes a global narrative arc before any writing begins, and the flexibility to select identical or varied settings and emotions across scenes accommodates different complexity levels (Fig. 1-a).

In story creation, the child composes six paragraphs (two per scene). The system generates eight contextually relevant keywords per paragraph via the ChatGPT API (GPT-3.5). This keyword mechanism embodies the core tension between structure and creative emergence identified in the formative research (Fan et al., 2024). *Four strongly associated keywords* align semantically with established narrative elements (e.g., "*crown*" or "*storybook*" when a child has described a princess reading in a garden). Four *weakly associated keywords* are semantically distant but thematically plausible (e.g., "*moonlight*" or "*treasure*" for the same context). Children with clear plans can select strongly associated keywords as vocabulary scaffolds, while those seeking creative stimulation can use weakly associated keywords for unexpected possibilities. Children can refresh keywords to regenerate based on updated story context. This keyword mechanism provides partial narrative information rather than complete AI-generated text, ensuring that children retain creative agency over story content while still receiving meaningful scaffolding. The system embeds scaffold questions at transition points and supports voice and text input (Fig. 1-b).

After each paragraph, the child can generate a comic-style illustration. ChatGPT extracts narrative elements from the text and formulates a prompt for Stable Diffusion to produce child-appropriate cartoon illustrations with consistent style tokens. Children can regenerate images until satisfied. The image generation function serves two purposes: it supports visual thinking by externalizing narrative content, and it opens additional storytelling possibilities when children notice discrepancies between their text and the generated image and choose to revise accordingly.

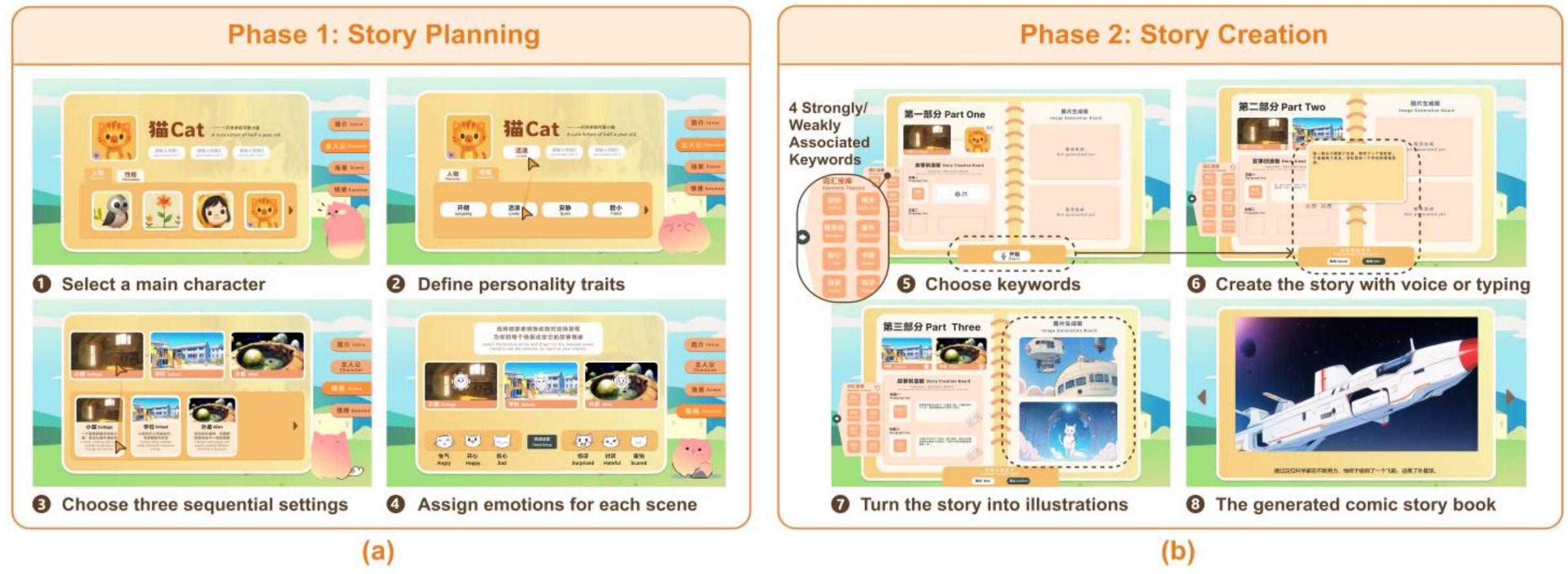

**Fig. 1** Overview of the *StoryPrompt* system. (a) Phase 1 (Story Planning): children select a character, define personality traits, choose three sequential scenes, and assign emotional arcs for each scene. (b) Phase 2 (Story Creation): children compose six story paragraphs using AI-generated keywords and generate comic-style illustrations after each paragraph.

## 3.2 Participants

Forty children from a public elementary school voluntarily participated in the study, recruited through school information flyers distributed to teachers and guardians. The sample comprised 30 boys and 10 girls, with an average age of 9.70 years (*SD* = 1.22, range 7–12)[4], spanning Grades 2 through 6. Following the grade classification used in the formative research and aligning with the developmental distinctions in prior work (Fan et al., 2024; L. Zhang, 2019), participants were divided into two grade groups: a lower-grade group (Grades 2–3, *N* = 22, mean age approximately 8.9 years) and a higher-grade group (Grades 4–6, *N* = 18, mean age approximately 11.1 years). All children had prior experience with computers or iPad applications. Most (*N* = 31, 77.5%) had previously explored AI products in some form, including AI drawing tools, AI writing assistants, or voice-based AI assistants. Children's general attitude toward AI products was positive, with a mean rating of 4.50 (*SD* = 0.64) on a 1–5 scale. Self-assessed creative writing ability was measured on a 5-point scale (1 = weak to 5 = excellent; distribution: 0% weak, 10% fair, 25% average, 37.5% good, 27.5% excellent). We measured self-assessed creative writing ability rather than storytelling ability because writing is the more familiar instructional category for elementary children in the Chinese school context, where "creative writing" is a standard curricular term.

Two participants (P36, Grade 3; P39, Grade 2) did not complete their storyboard stories due to time constraints, resulting in missing traditional storytelling data. P36 and P39 were excluded from all quantitative analyses (*N* = 38 for all reported statistics). Their observational and interview data from the AI condition were retained for qualitative analysis (*N* = 40).

Three elementary school teachers (with 1, 13, and 38 years of teaching experience respectively; two female, one male) observed sessions and participated in post-study interviews.

## 3.3 Study Procedure

Children were randomly assigned to two counterbalanced order groups (**Fig. 2**). We tested for order effects by comparing quality gains between children who completed the AI condition first (*N* = 18, *M* = 0.21, *SD* = 0.80) and those who completed the traditional condition first (*N* = 20, *M* = 0.32, *SD* = 0.52); no significant difference was found (Mann-Whitney $U = 185.5$, $p = .884$).

The study was conducted over two days in school classrooms, with an identical 30-minute storytelling lesson preceding each session (**Fig. 3**-a). After it, children selected their main character and scenes before beginning. In the AI condition, children first received a brief system tutorial (approximately five minutes) in which a facilitator walked through the interaction flow using a demonstration story. Children then worked individually on their stories using *StoryPrompt*, with a facilitator nearby to assist with technical issues (20–30 minutes per child). Voice recording was the primary input mode, supplemented by text editing (**Fig. 3**-b). In the Storyboard condition, children received paper sheets with six slots, markers, and reference images, sketching and writing or dictating stories that were recorded and transcribed (**Fig. 3**-c).

After each condition, children completed a semi-structured interview (10–15 minutes) on strategies, experience, and preferences. AI condition interviews additionally explored keyword and image usage. Eight trained facilitators followed standardized protocols. Facilitators were assigned to specific support roles, including AI system support (four facilitators), traditional storytelling support (two facilitators), and general coordination and documentation (two facilitators), rather than counterbalanced across conditions. Facilitator effects were not formally analyzed (**Fig. 2**).

[4] One participant's age (P33) was recorded incorrectly in the original dataset used for Fan et al. (2025) and has been corrected in the present analyses (from 8 to 12 years, consistent with the participant's Grade 6 enrollment). This correction results in slightly different sample descriptives (*M* = 9.70, *SD* = 1.22) compared to those reported in Fan et al. (2025; *M* = 9.64, *SD* = 1.18). All statistical analyses in the present manuscript use the corrected data.

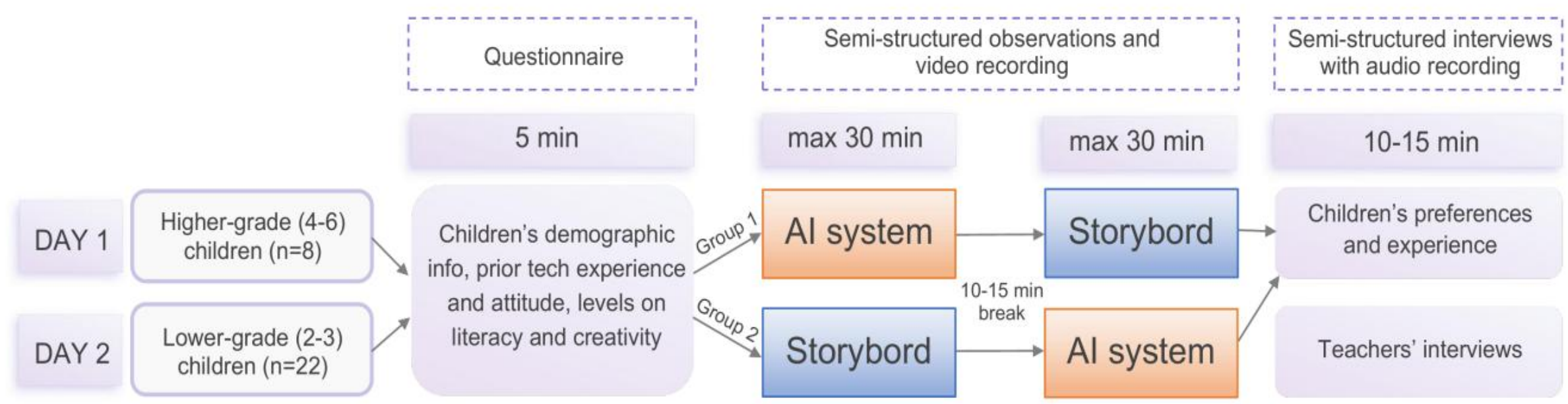

**Fig. 2** Overview of the study procedure: a counterbalanced within-subjects design conducted over two days. Each child completed a pre-study questionnaire and a storytelling instruction session, then participated in both the AI-assisted and traditional storyboard conditions (counterbalanced in order), followed by post-session interviews.

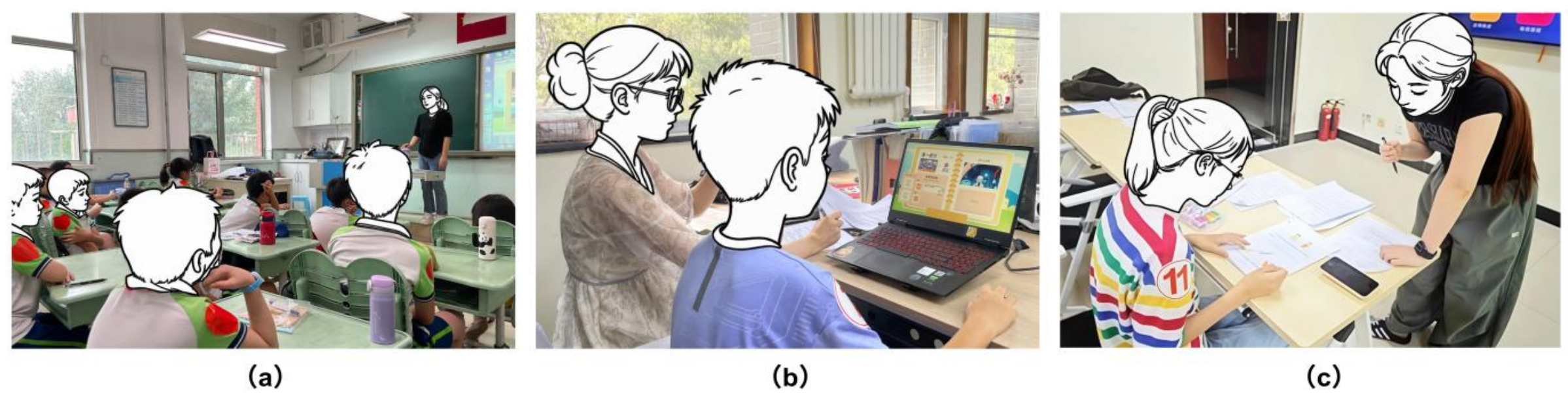

**Fig. 3** The three-stage experimental process. (a) A teacher-led storytelling instruction session for all participants. (b) The AI-assisted condition where a child uses the AI system individually. (c) The traditional storyboard condition where stories are sketched and written on paper.

## 3.4 Data Collection and Analysis

We combined quantitative statistical analyses with thematic analysis. Stories from both conditions were collected and anonymized. AI-condition story texts were captured through children's voice input and subsequent text edits, while storyboard texts were transcribed from children's oral presentations and notes. All texts were reviewed for completeness against recorded audio. All stories were created and rated in Chinese; text length refers to Chinese character count, and English story examples and interview excerpts were translated by the authors.

Five experts (one literacy teacher, one creative writing specialist, and three education/HCI researchers) rated all stories using CAT. Stories were rated on four quality dimensions, i.e., *creativity*, *richness*, *coherence*, and *narrative structure*, each on a 1-to-5 scale. Raters were blind to condition but were informed of each child's grade level so that quality could be evaluated relative to grade-appropriate expectations. Raters first rated 14 stories (18%) for calibration and discussed disagreements to finalize rating standards before independently rating the full set. Inter-rater reliability was good, with an average-measures ICC(2,k) of .832 (Baer & McKool, 2009). For each story, ratings were averaged per dimension, and composite quality was calculated as the four-dimension mean.

System logs captured keyword selections and image regeneration events for each paragraph. Strongly (or weakly) associated keyword count was calculated as the number of strongly (or weakly) associated keywords selected across the six story paragraphs. Image regeneration count was calculated as the total number of regeneration attempts across the six paragraphs, excluding the initial generation.

We report effect sizes and note trends ($0.05 < p \leq 0.10$) where relevant. **For RQ1 (differential quality gains)**, we used children's traditional quality as baseline and computed quality gain (AI minus traditional composite). Spearman correlations assessed how baseline was associated with gain; children were divided into thirds for Mann-Whitney $U$ comparisons. Cross-dimensional correlations tested dimension-selectivity.

**For RQ2 (engagement mechanisms across individual characteristics)**, three researchers conducted thematic analysis following Braun and Clarke's (2006) reflexive approach. First, all three independently reviewed the transcribed recordings, facilitator observation notes, and interview transcripts, generating initial codes for children's keyword selection behaviors and image generation interactions. Second, the researchers met to compare codes, resolve discrepancies through discussion, and iteratively group codes into candidate themes. Themes were refined until they coherently captured how children of different baseline levels and developmental stages engaged with AI features. Spearman correlations further tested whether age and baseline quality were associated with strongly associated keyword count.

**For RQ3 (engagement–outcome link)**, we computed Spearman correlations between each engagement metric (image regeneration count, strongly/weakly associated keyword count) and each AI story quality dimension (creativity, coherence, richness, narrative structure). Where engagement metrics correlated with baseline or age, partial correlations controlled for confounds.

Unless otherwise noted, quantitative analyses used the 38 children with complete paired stories. Qualitative case analyses also drew on AI-condition observations from the two children (P36, P39) without complete storyboard data.

### 3.5 Relationship to Prior Work and Novel Contributions

This manuscript reports a new secondary analysis of the same school-based evaluation dataset previously reported in Fan et al. (2025). The prior paper focused on the system design and the aggregate evaluation, reporting higher overall story quality, stronger creativity and richness, longer AI-assisted stories, no significant order effects, and broad keyword-use patterns. In contrast, the present manuscript asks a different question: for whom the AI-assisted condition was most beneficial, in which quality dimensions its effects operated, and how children's feature engagement related to outcomes. Accordingly, we introduce baseline-stratified and dimension-specific analyses, self-assessed ability analyses, engagement–outcome analyses, and mechanism-oriented qualitative re-analysis of logs, observations, interviews, and artifacts.

### 3.6 Ethical Considerations

The study protocol was approved by the Ethics Committee of the Communication University of China.The study was conducted in accordance with the ethical standards of the institutional research committee. Because the study involved elementary children interacting with GenAI, we implemented age-appropriate safeguards. The system used constrained prompts designed to generate age-appropriate keywords and images and to avoid violent or mature themes. Facilitators were present during all sessions to provide technical assistance and monitor children's interactions with the system. Written informed consent was obtained from all participating teachers and from the legal guardians of all child participants before the study. Child participants provided verbal assent after the researchers explained the study tasks, audio/video recording, voluntary participation, and their right to stop at any time. Children's personal identifiers were not included in the analyzed story artifacts or reported examples. Story artifacts and transcripts were anonymized, while audio/video recordings and observational data were stored securely and accessed only by the research team. Children and guardians provided informed consent/assent, and children were told they could stop at any time.

## 4. Results

We present results on the differential quality gains across individual characteristics (RQ1), individual behavioral patterns with AI assistance (RQ2), and the potential relationships between the two (RQ3).

### 4.1 Differential Quality Gains Across Individual Characteristics

The aggregate quality advantage masked substantial variation: 27 of 38 children showed positive quality gains under AI while 11 showed negative gains. Three patterns emerged from closer examination.

**Baseline-linked convergence.** Children's traditional-condition storytelling quality, used as a proxy for baseline storytelling performance, was strongly and negatively associated with AI-minus-traditional quality gain, $r_s = -0.730$, $p < .001$. To examine this pattern more closely, we ranked children by traditional composite quality and divided them into terciles: bottom third ($N$ = 13, range: 1.35–2.90), middle third ($N$ = 12, range: 2.90–3.40), and top third ($N$ = 13, range: 3.50–4.65). Children in the bottom third gained an average of +0.808 points, with particularly large gains in creativity (+1.292) and richness (+1.369). In contrast, children in the top third showed a negative mean gain (−0.292), with decreases concentrated in coherence (−0.431) and narrative structure (−0.558). The gain difference between the bottom and top thirds was significant, Mann-Whitney $U$ = 162.5, $p$ < .001 (**Fig. 4**).

This pattern should be interpreted cautiously because the traditional-condition score is also part of the AI-minus-traditional gain score. Two observations nevertheless support a substantive interpretation. First, the top third averaged approximately 3.8 on the 1–5 scale, so their decreases were not simply a ceiling effect; they were directional and concentrated in specific structural dimensions. Second, order effects were not evident: children who completed the AI condition first and those who completed the traditional condition first showed similar gains ($M$ = 0.21 vs. 0.32). No significant difference was found (Mann-Whitney $U$ = 185.5, $p$ = .884).

Self-assessed creative writing ability provided secondary convergent evidence. It was negatively associated with quality gain, $r_s$ = −.342, $p$ = .036, suggesting that children who rated their own writing ability lower tended to gain more under the AI-assisted condition. Because this measure was self-reported, ordinal, and moderately related to expert-rated baseline quality, we treat it as a complementary individual-difference indicator rather than a primary explanatory variable.

**Robustness checks.** To address possible mathematical coupling between baseline and gain, we conducted three sensitivity analyses. First, the negative baseline–gain association was significant within both order groups: AI-first, $r_s$ = −.886, $p$ < .001, $N$ = 18; traditional-first, $r_s$ = −.597, $p$ = .005, $N$ = 20. Second, cross-rater split analyses reduced shared rating noise by using one subset of raters to estimate traditional baseline and the remaining raters to estimate gain. Across all 10 possible 2/3 rater splits, the negative association remained significant, median $r_s$ = −.629, range [−.724, −.518], all $p$ < .01. Third, as a non-change-score check, we directly regressed AI-condition quality on traditional-condition quality. The slope was $\beta$ = 0.241, $SE$ = 0.107, 95% CI [0.031, 0.451], $R^2$ = .123, well below the identity-line slope of 1. This suggests that AI-condition scores were compressed relative to traditional-condition scores. Together, these checks indicate that the convergence pattern was not solely driven by order effects or shared rater-level measurement noise, although change-score coupling and regression-to-the-mean concerns cannot be fully eliminated without an independent pretest.

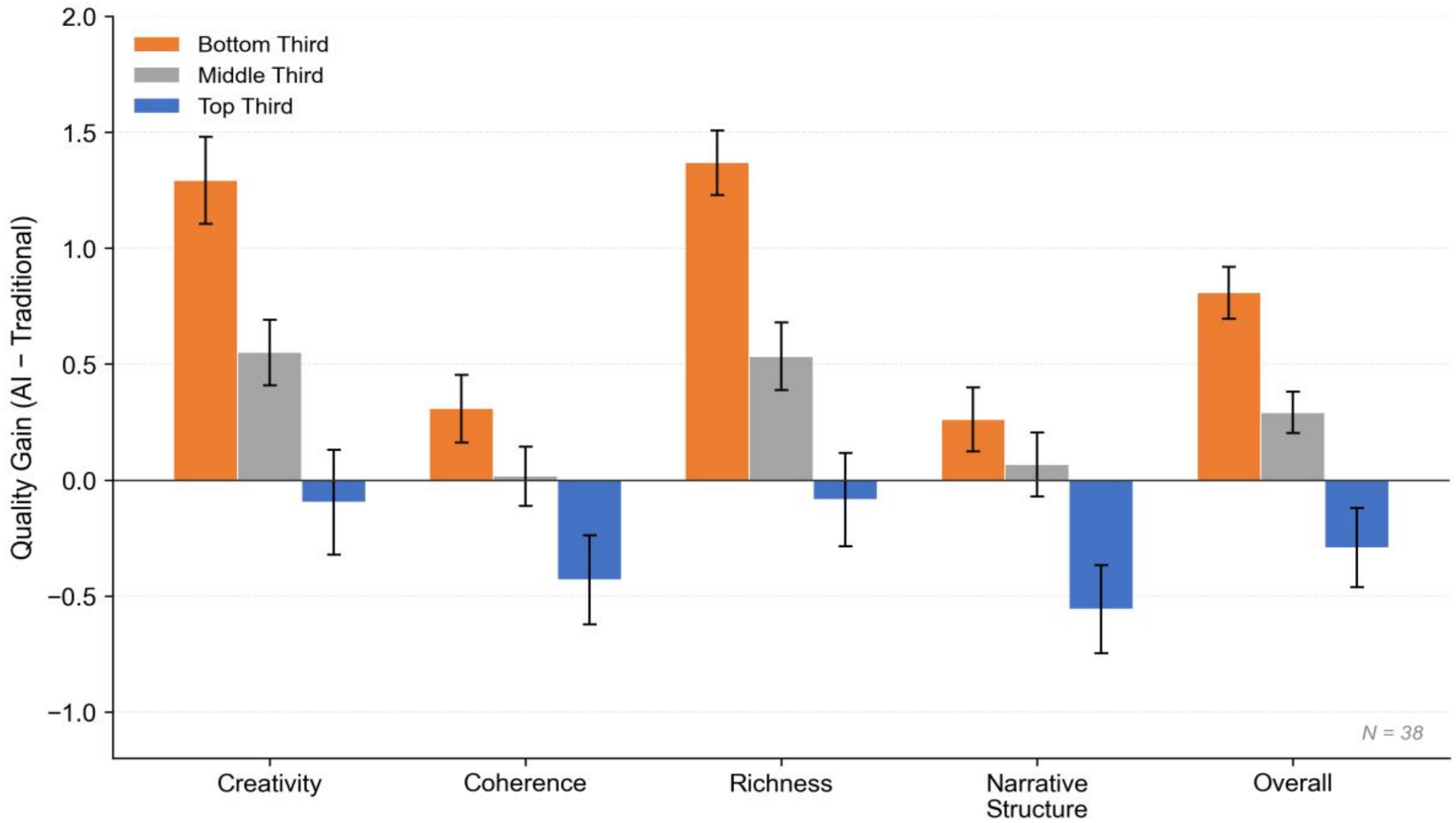

**Fig. 4** Quality gain (AI minus traditional) by performance tercile and evaluation dimension. Lower-performing children (orange) show substantial gains across all dimensions, particularly in creativity and richness. Higher-performing children (blue) remain

largely unchanged in creativity and richness but show notable decreases in coherence and narrative structure. Error bars indicate standard errors.

**Dimension-selective outcomes.** Cross-dimensional correlations showed that the AI-assisted condition was associated with different patterns across story-quality dimensions. AI coherence and narrative structure were significantly associated with all four traditional-condition dimensions, all $r_s > .39$, all $p < .05$, suggesting that these structural qualities remained closely tied to children's baseline storytelling competence. In contrast, no traditional-condition dimension was significantly associated with AI creativity or AI richness, all $p > .05$, suggesting that creativity and elaborative content were less constrained by baseline performance in the AI-assisted condition (**Fig. 5**).

This dimension-selective pattern also appeared in grade-group comparisons. Under the AI-assisted condition, lower-grade and higher-grade children achieved comparable age-normed richness ratings, $M = 3.547$ vs. 3.600, $p = .883$, whereas a trend toward a higher-grade advantage appeared under traditional conditions, $M = 2.726$ vs. 3.205, $p = .086$. By contrast, the age-normed narrative structure gap persisted under AI assistance, $M = 3.105$ vs. 3.558, $p = .010$. This pattern is consistent with the interpretation that AI-generated keywords and images supported content generation more directly than higher-order structural coordination.

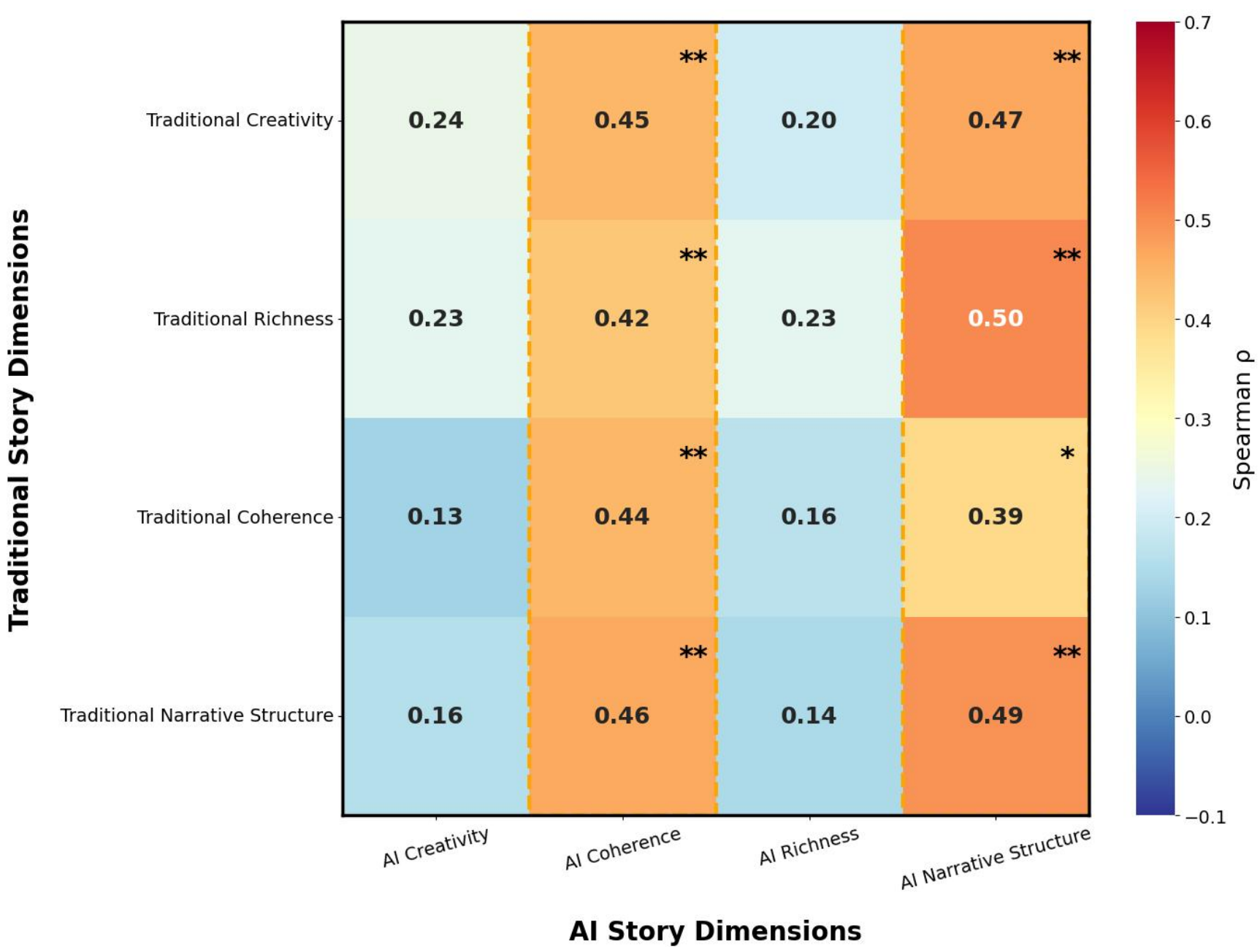

**Fig. 5** Cross-dimensional Spearman correlation matrix between traditional-condition and AI-condition story-quality dimensions ($N = 38$). Traditional-condition dimensions were significantly associated with AI coherence and AI narrative structure, whereas none was significantly associated with AI creativity or AI richness. *$p < .05$, **$p < .01$.

**Text length and score compression.** The AI-assisted condition also changed the relationship between text length and story quality. In the traditional condition, Chinese character count was strongly associated with composite quality, $r_s = .895$, $p < .001$. Under AI assistance, this association remained significant but weakened, $r_s = .634$, $p < .001$ (**Fig. 6**). The developmental text-length gap also narrowed: under traditional conditions, lower-grade children produced shorter stories than higher-grade children, $M = 152$ vs. 208 characters; under AI assistance, this gap effectively disappeared, $M = 240$ vs. 259, $p = .759$. Importantly, all narrative text in the AI condition was composed by children themselves; the system contributed six keywords and six illustrations per story, but no narrative prose. Because expert ratings may partially reflect the amount of narrative content available for evaluation, we interpret the text length–quality association descriptively rather than as evidence that length and quality are fully separable constructs.

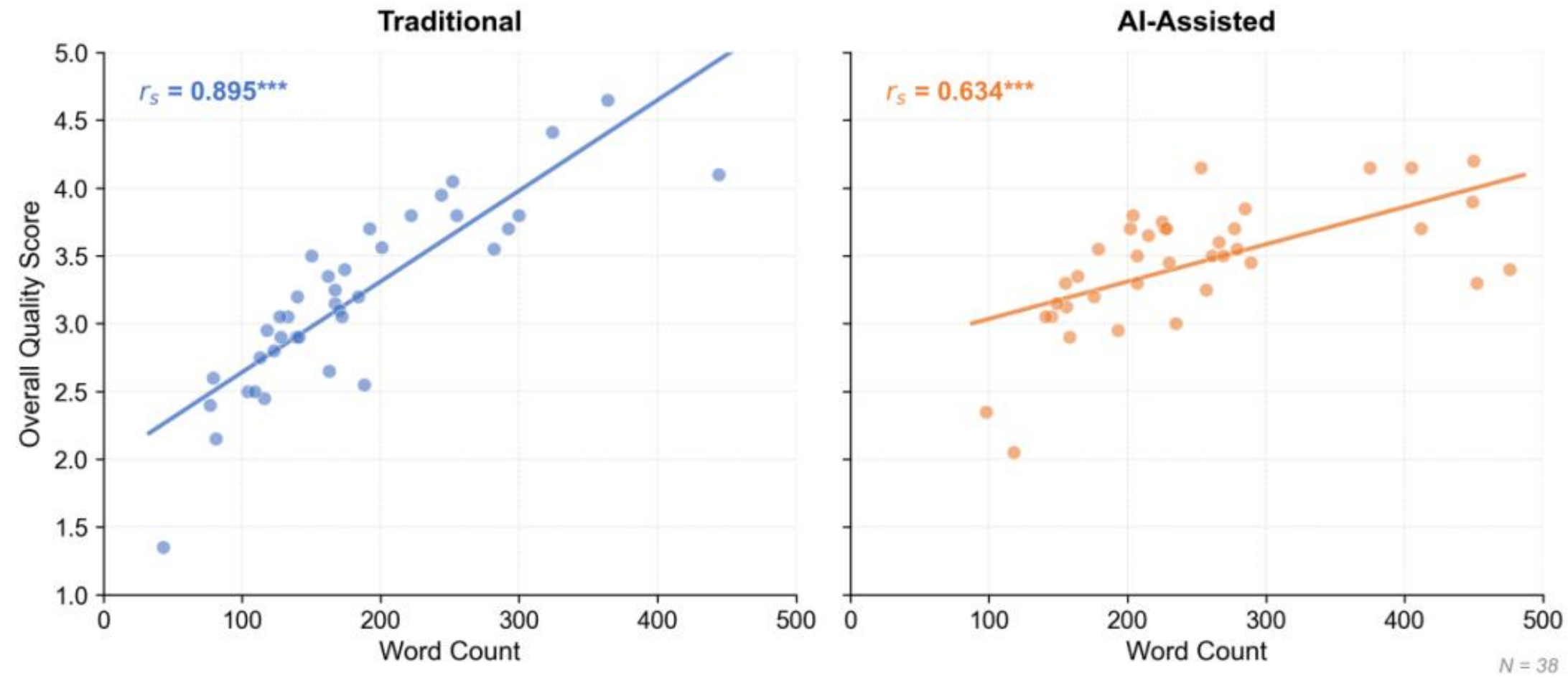

**Fig. 6** Association between text length, measured as Chinese character count, and overall story quality in the traditional and AI-assisted conditions ($N$ = 38). The strong association in the traditional condition weakened under AI assistance, suggesting that story quality was less tightly tied to output volume. ***$p$ < .001.

Taken together, these patterns are consistent with a floor-raising convergence pattern (**Fig. 7**). As a descriptive index of score compression, the median-split quality gap narrowed by 83.5%, from 1.028 points in the traditional condition to 0.170 points under AI assistance. This percentage should not be interpreted as an independent causal estimate of the system's equalizing effect (**Fig. 7**-a). AI-condition scores also showed descriptively smaller dispersion than traditional-condition scores for both overall quality, variance ratio = 2.12×, and creativity, ratio = 2.27×. Because these variance estimates come from paired observations of the same children, we treat them as descriptive evidence of score compression rather than formal inferential evidence (**Fig. 7**-b). Overall, the AI-assisted condition appeared to narrow performance gaps in generative dimensions, especially creativity and richness, while structural dimensions such as coherence and narrative structure remained harder to equalize.

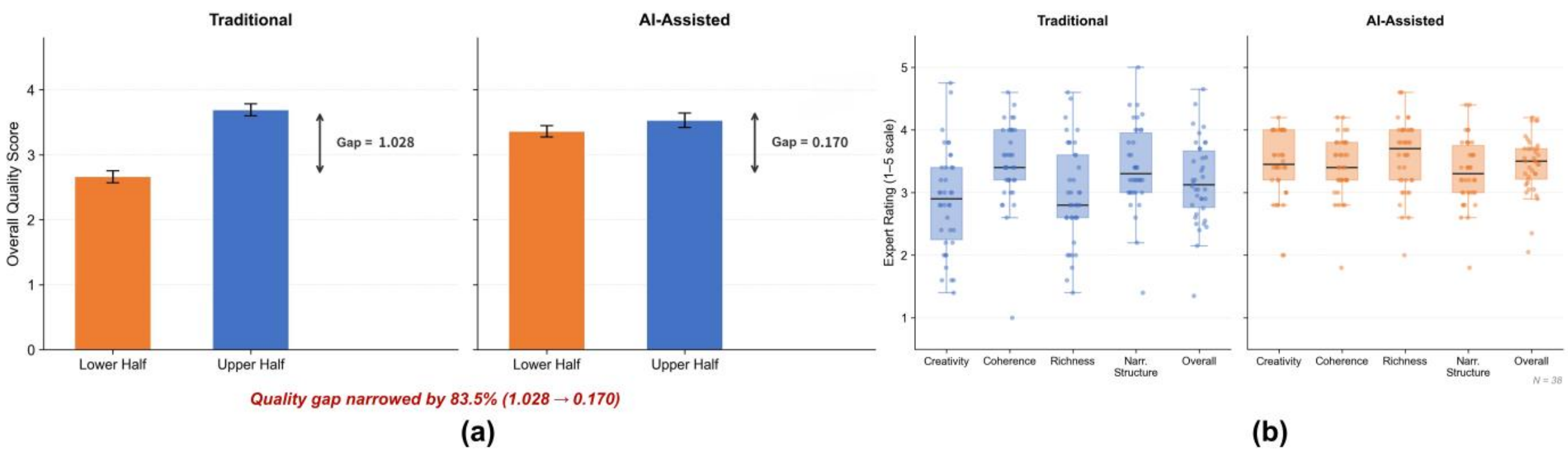

**Fig. 7** Floor-raising convergence pattern in story quality. (a) The median-split quality gap narrowed from 1.028 points in the traditional condition to 0.170 points under AI assistance, a descriptive reduction of 83.5%. (b) AI-condition scores also showed descriptively smaller dispersion in overall quality and creativity.

## 4.2 How Individual Characteristics Shape AI-Feature Engagement

Our prior work identified three keyword engagement modes (exploratory, goal-directed with openness, and plan-persistent[5]) and two image engagement orientations (productive alignment and reactive accommodation) (Fan et al.,

[5] Our prior work identified three keyword engagement modes termed *accepting*, *serendipitously incorporating*, and *strategically searching*, which we here shorten to *exploratory*, *goal-directed with openness*, and *plan-persistent* to more precisely capture the orientational distinction.

2025). Here, we examine how children's individual characteristics related to their engagement with the system's core AI features.

### *4.2.1 Baseline Performance and Differential Engagement Mechanisms*

Qualitative analysis revealed that children's baseline storytelling performance, that is, their within-age relative ability as assessed by experts, shaped not only how much they benefited from AI, but also *how* they engaged with both AI keyword and image features.

**For lower-performing children**, two distinct mechanisms emerged. The first was *compensatory scaffolding*, wherein children lacked narrative ideas to express and AI provided missing creative resources. P2 (Grade 3, traditional quality 1.35) exemplified this pattern. He adopted an exploratory approach, performing zero keyword refreshes, accepting every keyword, and building his entire story around them. He described his approach as "*making up the story step by step as I go*," with keywords having a "*big*" influence on plot development. His quality improved from 1.35 to 3.00 (+1.65, the largest gain in the sample) (**Fig. 8**). For these children, keyword engagement was not a complement to an existing narrative plan but a substitute for one, and AI-generated images served as confirmatory visual anchors rather than creative stimuli.

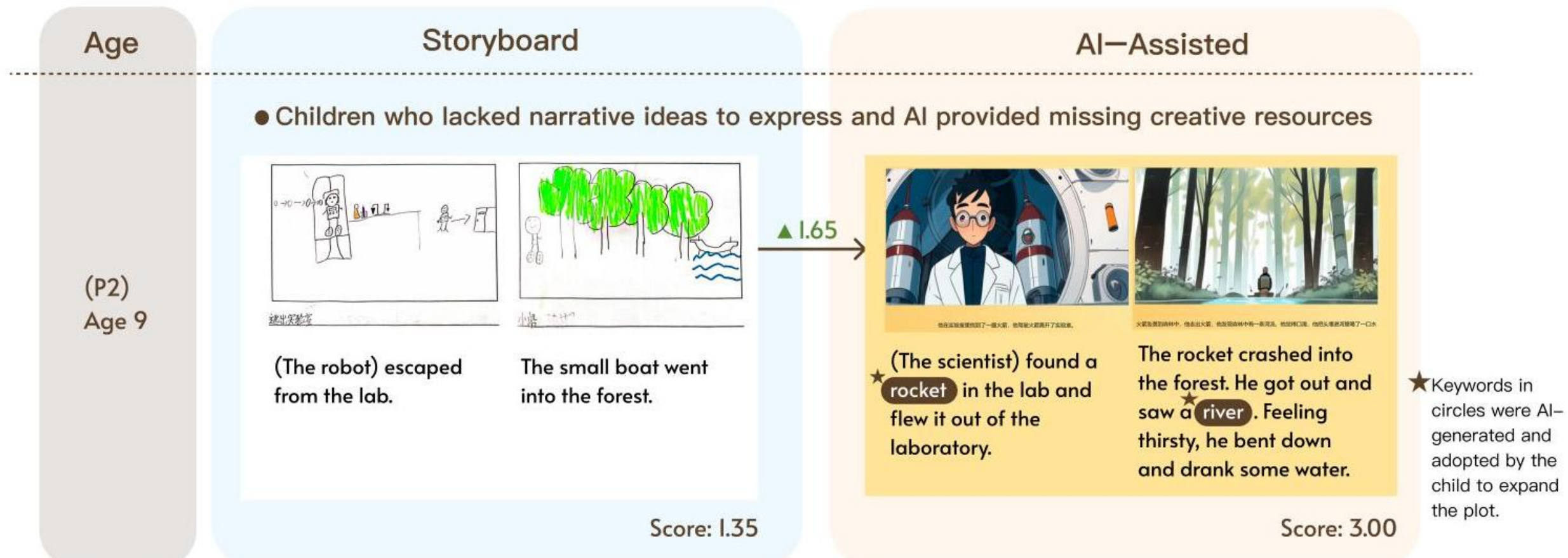

**Fig. 8** Compensatory scaffolding mechanism for a low-performing child showing narrative quality improvement (+1.65) through the adoption of AI-generated keywords (circled) as creative anchors.

The second was *restorative release*, in which the AI-assisted condition appeared to reduce non-narrative burdens, either execution barriers or cognitive overload, that had suppressed children's storytelling below their observable storytelling capacity. This mechanism operated through two related channels. The first channel involved *execution barrier removal*. In the traditional condition, children prepared their stories through hand-drawn storyboards with written notes before oral narration. Some children possessed narrative ideas but lacked the drawing and writing skills to construct an effective storyboard, resulting in poor visual and textual scaffolding for their subsequent narration. P12 (Grade 3, traditional quality 2.75) rated his own creativity and writing ability as 5/5, yet his traditional storyboard featured poorly executed drawings and limited written planning, and expert evaluators rated his resulting story low on coherence. Under AI (quality 3.65, +0.90), AI-generated images provided high-quality visual anchors that replaced his inadequate drawings, and AI-generated keywords offered narrative scaffolding that compensated for weak written planning. His AI story was recorded as "logically coherent," indicating that his narrative capacity had been present but trapped behind the storyboard preparation demands of the traditional format. The second channel involved *cognitive resource reallocation*. Unlike children in the first channel who lacked drawing and writing skills, these children appeared to over-invest in drawing. Several children invested substantially in elaborate, colored hand-drawn illustrations during the traditional storyboard phase, producing visually impressive preparation materials yet showing suppressed narrative quality, suggesting that the cognitive resources devoted to detailed visual production competed with those available for narrative composition. P31 (Grade 5, traditional text length only 81 Chinese characters) exemplified this most strikingly, producing elaborate drawings that appeared to consume the cognitive resources needed for storytelling. Under AI, her text length increased to 476 Chinese characters (+395) and she performed zero image regenerations, accepting every AI image without revision and redirecting all freed cognitive resources to narration. P9 and P5, both age 9, showed a similar pattern with elaborate traditional drawings but low story scores (2.55 and 2.45 respectively), improving substantially under AI (+1.15 and +1.00). This channel was moderated by cognitive maturity. Older children who drew equally elaborately (e.g., P33 and P38, both Grade 6) maintained high traditional quality (4.05 and 3.95) despite their drawing investment, suggesting their cognitive capacity was sufficient to sustain both activities simultaneously. This moderating role of cognitive maturity distinguishes the resource reallocation channel from the execution barrier channel, where the limitation is skill-based rather than capacity-based. Children benefiting from either restorative release channel typically engaged productively with both AI features, using keywords to anchor ideas they already had and accepting AI images as adequate visual representations, freeing attention for narrative elaboration (**Fig. 9**).

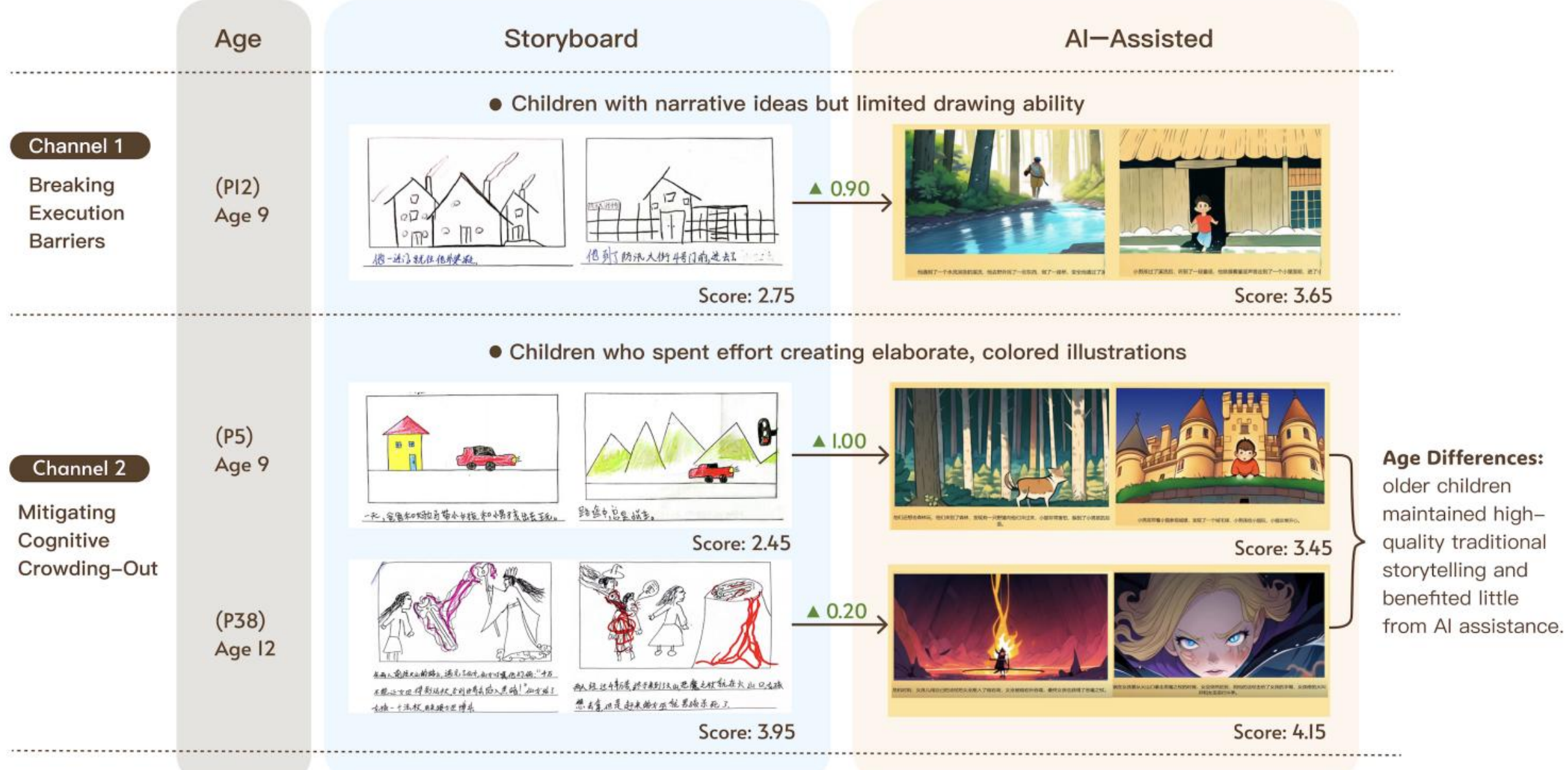

**Fig. 9** Restorative release mechanism via Channel 1 (execution barrier removal, illustrated by P12) and Channel 2 (cognitive resource reallocation, illustrated by P5). P38 serves as a developmental contrast case: older children with sufficient cognitive capacity to sustain both drawing and narration simultaneously maintained high traditional quality and gained minimally from AI assistance.

A boundary condition for both lower-end mechanisms also emerged. P10 (Grade 3, AI quality 2.35) showed systematic disengagement from the system's affordances, adopting a plan-persistent approach (repeatedly refreshing but refusing to incorporate unexpected keywords) and performing zero image regenerations. His facilitator noted that he "*basically selects school-related objects*" matching his predetermined story. His case suggests that AI scaffolding can compensate for expression barriers, ideational limitations, and cognitive overload, but not for disengagement from the creative interaction itself.

**For higher-performing children**, constraint patterns emerged that reflect misalignment between the child's capabilities and the system's uniform scaffolding. The most prominent was *scaffold-autonomy interference*. P25 (Grade 4, traditional quality 4.65, the highest in the sample) dropped to 3.30 (−1.35) because her autonomous narrative approach clashed with the system's structured scaffolding. P25 resisted the AI keyword and image scaffolding itself, asking with frustration: "*Can I type my own keywords?!*" For P25, the system's scaffolding features, designed to support children who need them, functioned as constraints that prevented her from deploying the autonomous creative approach that made her traditional story excel.

A distinct but related phenomenon was *visual-control interference*. P1 (Grade 3, traditional quality 3.80) performed 30 image regenerations (22 in the final paragraph alone), pursuing visual accuracy to the point of crowding out narrative construction. Unlike P25, who resisted the scaffolding itself, P1 over-engaged with a single feature, converting image regeneration from a support tool into a cognitive sink. While the underlying mechanisms differ, both cases share a common outcome: system features designed as supports functioned as constraints for these children.

Together, these cases suggest that AI scaffolding benefits are not automatic but depend on the alignment between the system's uniform design and each child's individual capabilities and engagement stance.

*4.2.2 Developmental Stage and Keyword Engagement Strategy*

Independent of baseline performance, developmental stage showed a trend-level relationship with the semantic distance of selected keywords, but not with quality gain or engagement orientation. Semantic distance refers to how closely a keyword aligned with the child's current narrative context, operationalized as strongly versus weakly associated keywords. Engagement orientation, by contrast, refers to the child's stance toward incorporating novel information into a preconceived plan: exploratory, goal-directed with openness, or plan-persistent.

Age was not significantly associated with AI-minus-traditional quality gain, $r_s = -.140$, $p = .400$, suggesting that developmental stage shaped engagement style more than gain magnitude in the present sample. A trend-level positive correlation emerged between age and the strongly associated keyword count, $r_s = .299$, $p = .068$, while baseline storytelling performance showed no such relationship, $r_s = .005$, $p = .977$. Because baseline storytelling performance was assessed relative to age-appropriate expectations, the absence of a baseline–keyword correlation suggests that semantic distance preference was not primarily driven by within-age ability differences. Instead, the trend-level age correlation may reflect a developmental tendency, possibly related to older children's greater use of goal-directed planning.

Qualitative cases suggested that younger children more often used weakly associated keywords as creative triggers. P2 (Grade 3) selected five of six weakly associated keywords and used each as a narrative turning point, building his story around words he had never sought. P5 (Grade 3) similarly used weakly associated keywords to discover story directions. Older children, by contrast, more often selected strongly associated keywords aligned with their emerging narrative context. P21 (Grade 4) actively searched for specific words like "*witch*" to match an intended scene, and P40 (Grade 5) described his process as "*I choose keywords based on the plot, first thinking about the plot.*"

This developmental shift in which type of keyword children preferred should not be equated with a shift in how open they were to novel information. A child selecting a strongly associated keyword may still be exploratory (willing to let the keyword redirect the story), and a child selecting a weakly associated keyword may still be plan-persistent (forcing the surprising word into a predetermined slot). The three engagement orientations we observed appeared across age groups rather than clustering by developmental stage. For example, P5 (Grade 3) embraced an unexpected AI image that depicted his intended villain as adorable and reversed the character's role, exemplifying exploratory openness regardless of keyword type. P21 (Grade 4), although primarily goal-directed, remained open to alternatives when his intended search proved unproductive, noticing "*forest elf*" and introducing a new character. P40 (Grade 5) articulated this dual orientation explicitly: "*first thinking about the plot, then occasionally a keyword like 'magic wand' would spark new imagination.*" In contrast, P10 (Grade 3), whose plan-persistent disengagement was discussed before, exhibited this orientation at the same developmental stage as the exploratory P2 and P5, refusing to incorporate any keyword that deviated from his predetermined plan. These cases indicate that engagement orientation is shaped by factors beyond developmental stage, possibly prior planning habits, personality, or the strength of a preconceived narrative goal, rather than by age alone.

Neither keyword semantic type independently predicted quality outcomes (see Section 4.3), suggesting that the developmental shift in semantic preference reflects a change in creative process rather than a difference in creative effectiveness. Whether engagement orientation independently predicts quality outcomes is a question that merits investigation with larger samples.

## 4.3 How Engagement Strategies Relate to Quality Outcomes

Given the differential engagement mechanisms identified above, we examined whether two specific engagement strategies, AI image regeneration count and AI keyword type (strongly or weakly associated), were associated with storytelling quality outcomes, and if so, in which dimensions.

Image regeneration count showed a dimension-selective correlation pattern. It was significantly positively correlated with coherence ($r_s$ = .375, $p$ = .021) and narrative structure ($r_s$ = .398, $p$ = .013) but not with creativity ($r_s$ = .169, $p$ = .312) or richness ($r_s$ = .182, $p$ = .274) (**Fig. 10**). That is, children who regenerated images more frequently produced stories with better structural quality, while regeneration bore no relationship to the content dimensions that AI keywords and images directly scaffolded. This pattern complements the dimension-selective scaffolding effect reported: AI-generated content elevated creativity and richness across ability levels, while structural quality remained tied to individual competence, and image regeneration may mark a candidate pathway for structural reflection through which some children achieved better structural outcomes.

However, this association requires cautious interpretation. Image regeneration count was not significantly correlated with age ($r_s = -.002$, $p$ = .990) but showed a trend-level association with baseline storytelling performance ($r_s$ = .318, $p$ = .052). Given this pattern, we conducted partial Spearman correlations controlling for baseline quality. The associations with coherence ($r_s$ = .266, $p$ = .107) and narrative structure ($r_s$ = .290, $p$ = .078) were attenuated to non-significance, suggesting that part of the zero-order association overlapped with baseline storytelling performance. The narrative-structure association remained directionally positive and trend-level after baseline control, but the present data cannot confirm that image regeneration independently contributes to structural quality beyond what baseline performance already predicts.

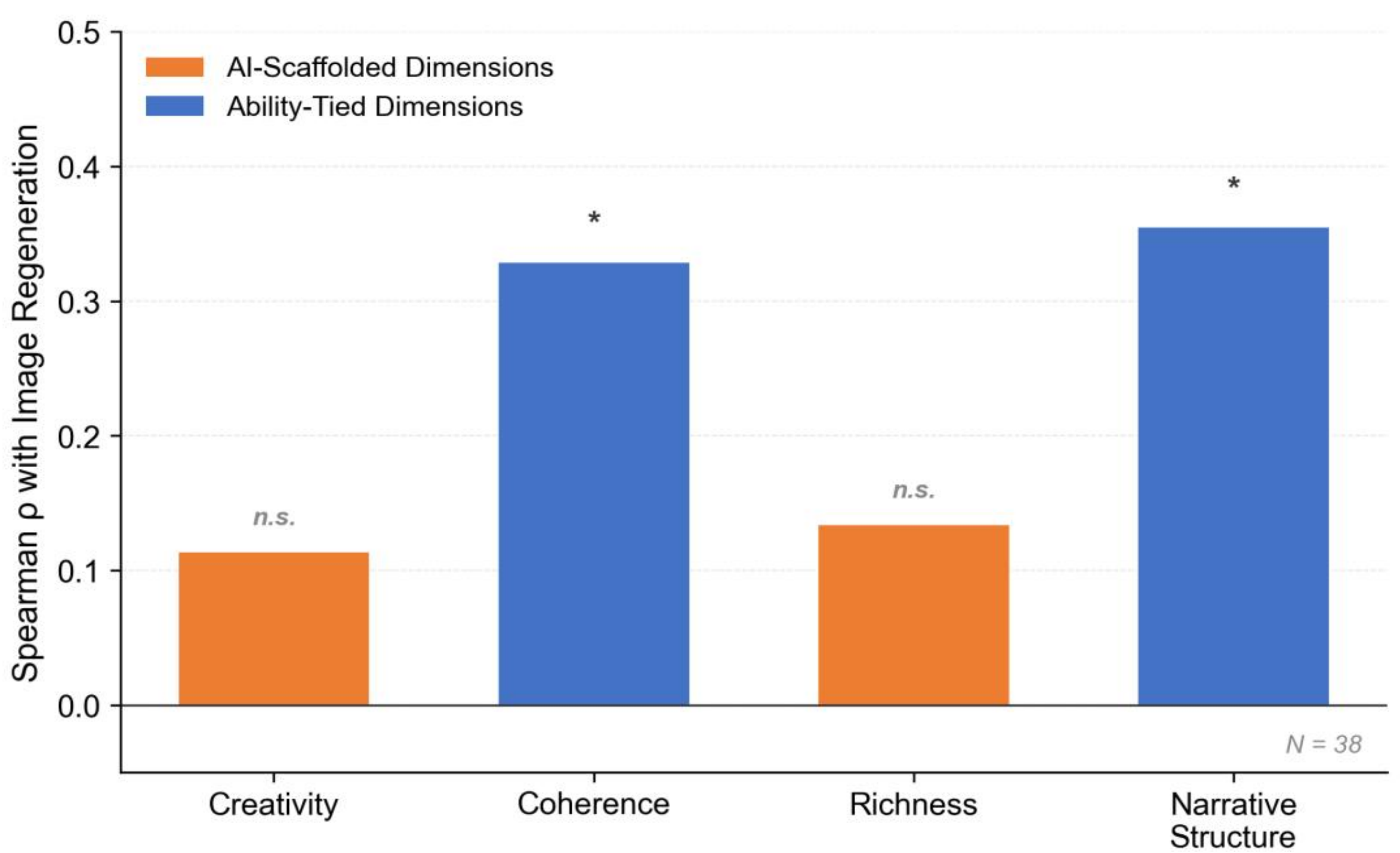

**Fig. 10** Zero-order Spearman correlations between image regeneration count and four AI storytelling quality dimensions ($N$ = 38). Image regeneration was significantly associated with the structural dimensions, coherence and narrative structure, but not with the content dimensions, creativity and richness. Baseline-controlled partial correlations were attenuated to non-significance, so these associations should be interpreted as exploratory. *$p$ < .05.

Keyword type, by contrast, showed no significant association with AI-condition quality outcomes. Neither the strongly associated keyword count nor the weakly associated keyword count was significantly associated with any

AI quality dimension (all $p > .05$). This null finding held regardless of age controls, despite the trend-level age-keyword correlation ($r_s = .299$, $p = .068$). The strongly associated keyword count was also unrelated to baseline storytelling performance ($r_s = .005$, $p = .977$). These results suggest that developmental stage shapes which type of keywords children prefer (a shift in semantic distance preference) but that this preference does not translate into quality differences.

Taken together, among the engagement behaviors examined, image regeneration was the only examined engagement metric showing positive zero-order associations with quality outcomes, and this association is confined to structural dimensions. The null finding for keyword type suggests that *how* children engage with AI features (their engagement orientation, as discussed in Section 4.2.2) may matter more for quality outcomes than *which* type of keywords they select, though the present sample was not powered to test engagement orientation as a quantitative predictor.

# 5. Discussion

In this section, we situate these findings within the broader literature on AI's equalizing effects on creativity, and discuss design implications and challenges for adaptive multimodal AI storytelling systems serving children across individual differences.

## 5.1 Interpreting the Convergence Pattern in Children's AI-Assisted Storytelling

Our most prominent finding, that the AI-assisted condition was associated with substantially larger gains for lower-performing children, while higher-performing children gained little or experienced selective decreases, converges with the equalizing pattern documented in adult creative writing (Doshi & Hauser, 2024) and professional productivity contexts (Brynjolfsson et al., 2025; Dell'Acqua et al., 2023). The strong negative baseline–gain correlation, gap narrowing, and variance compression we observed parallel Doshi and Hauser's finding that AI brought lower-creativity adult writers to parity with their higher-creativity peers. Children with high self-assessed creative writing ability showed no significant gain ($p = .216$), providing secondary convergent evidence that larger gains were concentrated among children with lower baseline or self-perceived ability. At the surface level, then, children's AI-assisted storytelling exhibits the same equalizing signature as adult AI-assisted creative work. However, the evidence suggests three ways in which this signature differs from adult AI-assisted creativity findings, which together refine the resource-provision framework that has dominated the adult literature.

First, the convergence in our data is a two-way process rather than a single-ended lift. Adult equalization studies have focused almost exclusively on how AI raises the performance of lower-skilled users (Brynjolfsson et al., 2025; Doshi & Hauser, 2024); the upper end is typically described as unaffected or modestly improved. In our data, by contrast, higher-performing children sometimes declined under AI, and this upper-end movement contributed meaningfully to the gap reduction. Our qualitative analysis traced these declines to two related constraint patterns. P25 illustrates *scaffold-autonomy interference*: she resisted the structured keyword and image scaffolding itself, declining from 4.65 to 3.30. P1 illustrates *visual-control interference*: he converted image regeneration from a support feature into a cognitive sink by pursuing visual control to the point of crowding out narrative construction. These two cases represent distinct forms of the same underlying mismatch (scaffold-autonomy interference in P25's case, visual-control interference in P1's) but share a common source. Scaffolding calibrated to children who need it functioned as a constraint for children who did not. Dell'Acqua et al. (2023) documented adult performance declines when tasks fell outside AI's capability frontier; our cases reveal a different source of decline, rooted not in the tool's capability boundary but in the uniformity of its scaffolding. This mechanism is theoretically important because resource-provision frameworks do not fully account for upper-end movement, yet upper-end movement is half of what generates a convergence pattern.

Second, the lower-end gains we observed arose from two qualitatively distinct mechanisms rather than a single resource-provision process. Adult studies share a common explanatory framework in which AI benefits lower-performing individuals by supplying resources they lack, whether creative ideas (Doshi & Hauser, 2024), tacit expert knowledge (Brynjolfsson et al., 2025), or relief from technical writing demands (Cen & Shakibaei, 2026). In each case, AI supplemented a deficit. Our analysis identified one mechanism that matches this framework directly and a second that does not. The first, *compensatory scaffolding*, corresponds closely to the adult resource-provision account. Children like P2 (traditional quality 1.35, AI gain +1.65) received new creative content, with AI-generated keywords functioning as narrative seeds that expanded their repertoire beyond what they produced in the traditional condition. The second, *restorative release*, reveals a process that has received limited attention in the adult literature, though Cen and Shakibaei's (2026) finding that AI relieved technical writing demands suggests a related process may operate in adult contexts as well. The developmental form we documented, however, involves barriers that have limited direct parallels in adult creative-writing tasks. Some children appeared to have narrative ideas that were obscured by non-narrative production burdens, whether literacy and fine-motor limitations (P12) or elaborate hand-drawing that consumed cognitive resources needed for narrative composition (P31, P9, P5). These cases suggest that low traditional-condition quality did not always reflect a lack of ideas or narrative skill; AI's contribution was not to supply missing resources but to reallocate cognitive resources from non-narrative production burdens back to narrative composition. Adult creative-writing tasks rarely involve the literacy, fine-motor, and hand-drawing barriers documented in P12's case, and the cognitive-resource form of the mechanism depends on developmental constraints specific to children. In line with Cognitive Load Theory (Sweller, 1988), restorative release appeared to operate selectively among children whose cognitive capacity was insufficient to sustain concurrent visual and narrative demands. Older children who drew equally elaborately (P33, P38, Grade 6) maintained high traditional quality and gained minimally from AI, suggesting that there was less suppressed potential to release in these cases. While these contrasting cases are suggestive, larger within-age-group samples would be needed to confirm this age-moderated boundary. This observation extends the cognitive resource reallocation proposal from visual art education (Bian et al., 2025) in two ways: we provide participant-level evidence suggesting how cognitive resource reallocation may be distinguished from other AI benefits, and we identify a developmental boundary condition on when the mechanism operates. The two lower-end mechanisms are not mutually exclusive and may operate simultaneously within the same child.

Third, the convergence is dimension-selective rather than uniform across quality dimensions. Adult equalization studies have largely treated creative output as a single construct, reporting composite quality gains without disaggregating which dimensions absorbed the benefit. Our data reveal a clear asymmetry: the AI-assisted condition was associated with less baseline-constrained creativity and richness across ability levels, while coherence and narrative structure remained associated with traditional-condition scores, with lower-grade children's structural quality still significantly below higher-grade children's even under AI ($p$ = .010). Adapting Dell'Acqua et al.'s (2023) "jagged technological frontier" concept, we can interpret this asymmetry as reflecting the boundary of the system's *scaffolding* frontier rather than its *capability* frontier per se. AI-generated keywords and images fell within the system's scaffolding frontier for content generation, directly supporting creativity and richness. By contrast, although the system provided a basic structural frame through scenes, emotions, and six story paragraphs, it did not adaptively scaffold the higher-order coordination needed to maintain temporal sequence, causal logic, and emotional arcs across the whole story. These structural demands therefore remained more dependent on children's own narrative coordination resources. Image regeneration count showed a tentative positive association with coherence and narrative structure, suggesting that iterative visual–textual comparison may serve as an incidental reflection cue for structural quality, though this association was largely attenuated when baseline performance was controlled and we treat it as a candidate pathway for future experimental validation rather than an established finding. To our knowledge, this dimension-selective pattern has not been directly examined in children's AI-assisted creative storytelling.

Taken together, these three theoretical refinements recast the convergence pattern as a two-way, heterogeneous, dimension-selective process rather than a single-ended, uniform resource-provision effect. The descriptive 83.5% gap narrowing appears to arise from the combined operation of two lower-end support mechanisms, compensatory scaffolding and restorative release; two upper-end constraint patterns, scaffold-autonomy interference and visual-control interference; and a dimension-specific scaffolding frontier that supported content generation more directly than structural coordination. This interpretation extends adult AI-equalization findings by showing that, in children's multimodal storytelling, convergence depends jointly on the child's baseline performance and the quality dimension being scaffolded.

## 5.2 Disentangling Baseline Performance and Developmental Stage

A second contribution of our analysis concerns how we decomposed "individual differences." Prior child–AI storytelling research has noted age-related variations in children's AI interactions, including differences in response elaboration across age groups (Liu et al., 2025) and developmental thresholds for narrative transfer after AI interaction (Ye et al., 2025), but has not systematically distinguished these developmental effects from within-age ability differences. This conflation matters because it leaves open whether observed age effects reflect developmental maturation per se or unobserved baseline performance variation correlated with age. Because our expert raters evaluated stories against grade-level expectations (i.e., a well-structured story from a Grade 2 child and a well-structured story from a Grade 5 child could both receive high ratings), baseline storytelling performance functions as a within-age relative measure. This, combined with disaggregated analysis by both dimensions, allows us to analytically distinguish the contributions of developmental stage and baseline performance.

The two dimensions contributed differently. Baseline storytelling performance was most closely associated with the convergence pattern. It was linked to the magnitude and direction of AI-minus-traditional gain and to which support mechanism or constraint pattern appeared most salient for a given child: compensatory scaffolding for children who lacked narrative content, restorative release for children whose performance appeared suppressed by execution barriers or cognitive resource competition, and upper-end constraint patterns, including scaffold-autonomy interference and visual-control interference, for higher-baseline children whose autonomous strategies or visual-control efforts clashed with the system's uniform scaffolding. Developmental stage, by contrast, showed no significant relationship with the magnitude of AI-minus-traditional gain and showed a trend-level relationship with the semantic distance of selected keywords. Exploratory quantitative trends and qualitative cases suggested that younger children more often used weakly associated keywords as creative triggers, whereas older children more often selected strongly associated keywords aligned with emerging plans. Importantly, this developmental shift concerned keyword semantic preference rather than engagement orientation. The three engagement orientations, i.e., exploratory, goal-directed with openness, and plan-persistent, appeared across age groups, and neither keyword semantic type independently predicted quality outcomes. The age-related shift therefore reflects a possible difference in creative process, namely how children approach keyword selection, rather than a demonstrated difference in story-quality outcomes.

This distinction clarifies how age effects should be interpreted in the child–AI storytelling literature. In our sample, developmental stage was more visible in how children engaged with AI features, whereas baseline performance was more closely tied to how much they benefited. Age effects documented in prior work (Liu et al., 2025; Ye et al., 2025) may index developmental process differences, unmeasured baseline performance differences, or both. These two patterns call for different responses: developmental process differences suggest age-appropriate interaction modalities, such as adjusting keyword presentation for younger versus older children; baseline performance differences suggest that scaffolding type and intensity should adapt to the child's operative constraint rather than to age alone.

## 5.3 From Mechanisms to GenAI Design Implications

The two lower-end support mechanisms and two upper-end constraint patterns identified above do not merely explain why the AI-assisted condition benefited some children and constrained others. They specify different design conditions under which scaffolding should be adapted. The design implication is therefore not simply to provide "more" or "less" support, but to infer the operative constraint and provide the corresponding type of scaffold.

*Restorative release* points to design opportunities for reducing non-narrative burdens. For children facing execution barriers, such as limited writing fluency or drawing skill, systems should provide alternative expressive modalities that bypass skill limitations, including voice-to-text input, simplified visual planning tools, or low-effort ways to externalize ideas. For children experiencing cognitive resource competition, systems should detect behavioral signals of over-investment in non-narrative production, such as extensive drawing time relative to narrative output, and redirect attention toward narrative composition. Such intervention should be calibrated to observable behavior rather than age alone, because our cases suggest that similar drawing investment may suppress narrative quality for some children but not for older or more coordinated children.

*Compensatory scaffolding* calls for adaptive ideation support. For children who lack a stable narrative plan, AI-generated keywords can function as narrative seeds rather than optional embellishments. A productive design response would be to help children transform these external prompts into self-directed story plans. For example, after a child repeatedly accepts keywords without an evident plan, the system could offer lightweight causal prompts, such as "*How did this object enter the story*?" or "*What problem could this keyword create*?" The goal is not merely to supply more keywords, but to scaffold the child's movement from prompt dependence toward independent story construction.

*Scaffold-autonomy interference* points to graduated scaffolding autonomy. P25's case shows that uniformly applied scaffolding can constrain children whose capabilities already exceed the system's default support. High-baseline children should therefore be able to bypass keyword constraints, type their own terms, reduce prompt count, or skip image generation when it does not serve their story plan. Children who benefit from full scaffolding can retain it, while children with stronger autonomous strategies can selectively withdraw from it. This is not merely a usability issue; it is a structural requirement for preserving creative voice at the upper performance range.

*Visual-control interference* suggests a different design response: guided image control rather than unrestricted regeneration. P1's case shows that image regeneration can become a cognitive sink when children pursue visual accuracy at the expense of narrative construction. At the same time, the image-regeneration results suggest a candidate opportunity: iterative visual–textual comparison may serve as a reflection cue for coherence and narrative structure, although this association was attenuated after controlling for baseline performance and should not be treated as an established mechanism. A useful design modification would be to ask children to specify what the image should show before generating it, or to explain what changed between the image and their intended story after regeneration. This would transform image generation from a matching task into a narrative planning and reflection activity, scaffolding structure through externalized planning rather than direct content provision.

Across these four design directions, engagement orientation remains a cross-cutting condition. A child who approaches AI suggestions exploratorily may use the same feature productively, while a plan-persistent child may reject or neutralize it. Thus, mechanism-contingent scaffolding should not only adapt feature intensity to baseline level, but also respond to how children are engaging with the scaffold in the moment. This aligns with Kupers et al.'s (2019) call for micro-level process scaffolding in children's creative technology interactions.

Together, these four mechanism-contingent design directions extend the conventional approach of calibrating support intensity to ability level. Ability-adaptive scaffolding is necessary but insufficient: children at the same baseline level may require different types of scaffolding depending on the operative constraint. A child whose low quality stems from cognitive resource competition needs drawing-burden reduction, not more keywords; a child whose low quality stems from ideational limitation needs adaptive ideation support, not image-control tools; a high-baseline child constrained by uniform prompts needs greater autonomy, not stronger scaffolding; and a child over-investing in image regeneration needs guided reflection, not unlimited visual iteration. The key design implication is that systems should either infer operative constraints from behavioral signals or help teachers identify them, and then deliver the corresponding scaffolding type rather than uniformly increasing or decreasing scaffolding intensity.

## 5.4 Tensions, Trade-offs, and the Limits of Scaffolding

The floor-raising convergence produced more equalized quality scores but also compressed the quality distribution. At the population level, lower-performing children's gains narrowed the between-child gap; at the individual level, some higher-performing children experienced quality decreases when the system's scaffolding mismatched their capabilities. This pattern resonates with Doshi and Hauser's (2024) finding that AI-assisted creative outputs can become more similar, although our evidence concerns quality-score compression rather than direct textual similarity. The broader tension is that floor-raising scaffolding may improve access to quality while also risking reduced distinctiveness for children who already have strong creative voices.

This tension is compounded by the ambiguity of what lower-end gains represent. As discussed in Section 5.1, the lower-end gains we observed arose from two support mechanisms operating in parallel: restorative release, in which the AI-assisted condition appeared to free storytelling capacity suppressed by execution barriers or cognitive resource competition, and compensatory scaffolding, in which AI-provided content resources supported story production beyond what children produced in the traditional storyboard condition. The present data do not allow a quantitative partition between the two, and individual children likely experienced both. This distinction matters for how educators interpret AI-assisted outcomes. A child whose AI-condition score substantially exceeds their traditional-condition score may have expressed existing capability freed from non-narrative burdens, may have extended their repertoire through AI-provided content, or both. Sustained creative development may therefore require practice that builds these capacities independently, not only conditions that compensate for their absence. Teachers may need to complement AI-assisted practice with self-directed creative expression to preserve individual voice development.

The boundary cases in our data further illuminate the limits of scaffolding. P25, the highest-performing child, experienced the largest quality decrease not because the system lacked content resources, but because she could not customize the core scaffolding mechanism. This inverts Han and Cai's (2023) concern that unconstrained AI generation risks reducing independent creative thinking: in our case, uniformly applied scaffolding appeared to constrain a strong storyteller who needed greater control over the scaffolding. This boundary motivates graduated scaffolding autonomy, as discussed in Section 5.3.

A further boundary concerns engagement orientation itself. AI scaffolding may help with expression barriers, ideational limitations, and cognitive overload, but it cannot substitute for generative engagement with the creative task. P10 exemplified this limit. Despite having access to the full range of AI features, his plan-persistent orientation led him to reject any keyword that deviated from his predetermined narrative, effectively neutralizing the system's scaffolding potential. This suggests that scaffolding effectiveness may depend partly on the child's willingness to engage generatively, an orientation that may require pedagogical cultivation. Consistent with Cen and Shakibaei's (2026) finding that AI-instructor interplay fostered greater learner independence than AI tools used in isolation, teachers may need to explicitly model exploratory engagement strategies rather than treating AI tools as self-sufficient scaffolds.

For classroom deployment, teachers should not treat GenAI storytelling tools as uniform scaffolds or self-sufficient tutors. Lower-baseline children may benefit from keyword and visual supports that lower expressive barriers, while higher-baseline children may need optional or bypassable scaffolds to preserve autonomy. The educational value of GenAI storytelling systems therefore lies not only in improving story products, but in enabling teacher-orchestrated, differentiated creative literacy support across diverse classroom populations.

# 6. Limitations

Several limitations should be noted. First, the sample comprised 40 children (38 with complete paired data) from a single school, limiting generalizability. The gender imbalance (30 boys, 10 girls) precluded robust gender-based analyses. Baseline storytelling performance was assessed by expert raters and therefore provides a more externally rated and standardized measure than self-report, but the age-normed rating procedure means that scores reflect within-age relative performance rather than absolute skill. Direct cross-age comparisons should therefore be interpreted cautiously. Grade-group comparisons should also be interpreted cautiously because lower- and higher-grade children were tested on different days, although both sessions followed the same procedure and instruction protocol. Self-assessed creative writing ability, used as a secondary measure, may not capture objective skill levels.

Second, the cross-sectional design cannot determine whether the observed convergence pattern persists over repeated use, diminishes as novelty fades, or translates into long-term storytelling development. The 20–30 minute storytelling window may also have been insufficient for higher-grade children to develop complex narratives, potentially underestimating the benefit of the AI-assisted condition for this group.

Third, although the within-subjects design controlled for between-child variability, traditional-condition quality served as a proxy baseline rather than an independent pretest. Counterbalancing and sensitivity checks reduced concerns about order effects and shared rater-level measurement noise, but change-score coupling and regression-to-the-mean concerns cannot be fully eliminated without an independent pretest.

Fourth, the AI-assisted condition bundled multiple forms of support, including AI-generated keywords, AI-generated images, structured story planning, voice/text input, and digital interaction. The present study therefore evaluates the system as an integrated GenAI-supported storytelling environment rather than isolating the independent causal effect of any single AI feature. Facilitators were also assigned to different support roles, and facilitator effects were not formally analyzed.

Finally, our quantitative analyses are correlational and cannot establish causal mechanisms. The support mechanisms, constraint patterns, and engagement orientations identified through qualitative analysis represent plausible interpretations requiring experimental validation. Because the feature–outcome and subgroup analyses were exploratory and involved multiple comparisons, p-values should be interpreted as pattern-identifying rather than confirmatory. We emphasize findings that converge across quantitative patterns and qualitative evidence.

# 7. Conclusion

We presented a mixed-methods within-subjects study of differential storytelling outcomes using a child-centric GenAI storytelling system with 40 elementary children, including 38 with complete paired stories for quantitative analyses. The AI-assisted condition was associated with a floor-raising convergence pattern: as a descriptive index of score compression, the median-split quality gap narrowed by 83.5%. Qualitative analysis suggested two lower-end support mechanisms, compensatory scaffolding and restorative release, and two upper-end constraint patterns, scaffold-autonomy interference and visual-control interference. This convergence was dimension-selective, with creativity and richness less constrained by baseline performance under AI assistance, while coherence and narrative structure remained more closely tied to children's baseline storytelling performance and narrative coordination resources. The findings also disentangle two individual-difference dimensions. Baseline storytelling performance was most closely associated with gain magnitude and with the support mechanism or constraint pattern that appeared most salient for each child. Developmental stage did not significantly predict gain magnitude, but exploratory trends and qualitative cases suggested age-related differences in keyword semantic preference. Image regeneration showed positive zero-order associations with coherence and narrative structure, though these associations were attenuated after baseline control and require larger-sample validation.

Overall, the study extends adult findings on AI equalization to children's multimodal storytelling by showing that convergence can be a two-way process involving both lower-end support and upper-end constraint. It also proposes mechanism-contingent scaffolding as a design principle for adaptive GenAI storytelling systems: rather than uniformly increasing or decreasing scaffolding intensity, such systems should identify operative constraints, whether through system inference, teacher orchestration, or both, and provide corresponding forms of support.

# Acknowledgements

This work was supported by the National Social Science Fund of China under Grant and the Fundamental Research Funds for the Central Universities under Grant. We would like to thank the teachers and students who participated in our study.

## Declarations of interest statement

The authors have no relevant financial or non-financial interests to disclose.

## Data Availability Statement

All the datasets can be accessed from https://figshare.com/s/1070741660bba1dfc219

# Figures

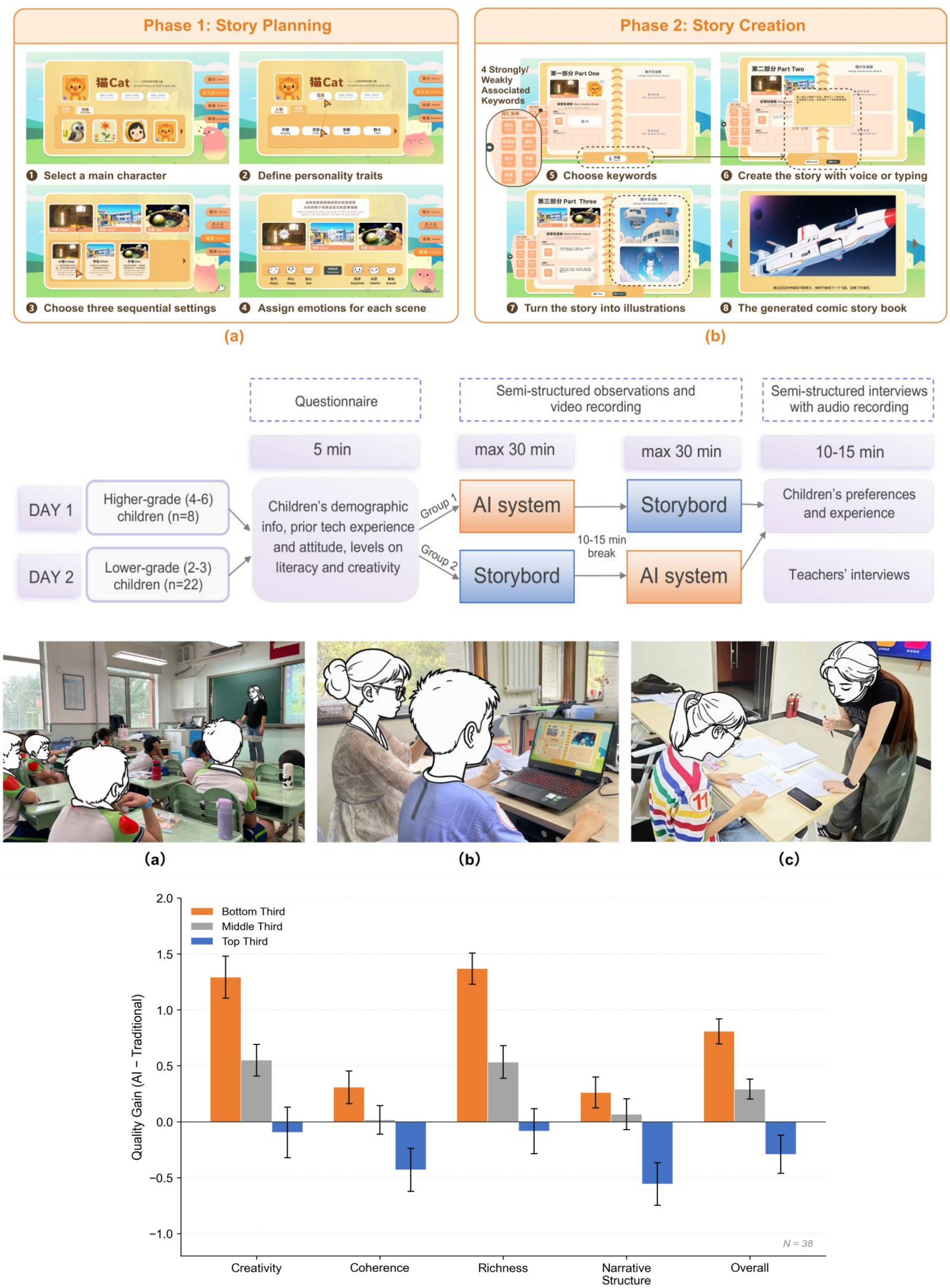

Phase 1: Story Planning
Phase 2: Story Creation
❶ Select a main character
❷ Define personality traits
❸ Choose three sequential settings
❹ Assign emotions for each scene
4 Strongly/ Weakly Associated Keywords
❺ Choose keywords
❻ Create the story with voice or typing
❼ Turn the story into illustrations
❽ The generated comic story book
(a)
(b)
Questionnaire
Semi-structured observations and video recording
Semi-structured interviews with audio recording
5 min
max 30 min
max 30 min
10-15 min
DAY 1
Higher-grade (4-6) children (n=8)
DAY 2
Lower-grade (2-3) children (n=22)
Children's demographic info, prior tech experience and attitude, levels on literacy and creativity
Group 1
Group 2
AI system
Storybord
Storybord
AI system
10-15 min break
Children's preferences and experience
Teachers' interviews
(a)
(b)
(c)
Bottom Third
Middle Third
Top Third
Quality Gain (AI − Traditional)
2.0
1.5
1.0
0.5
0.0
−0.5
−1.0
N = 38
Creativity
Coherence
Richness
Narrative Structure
Overall

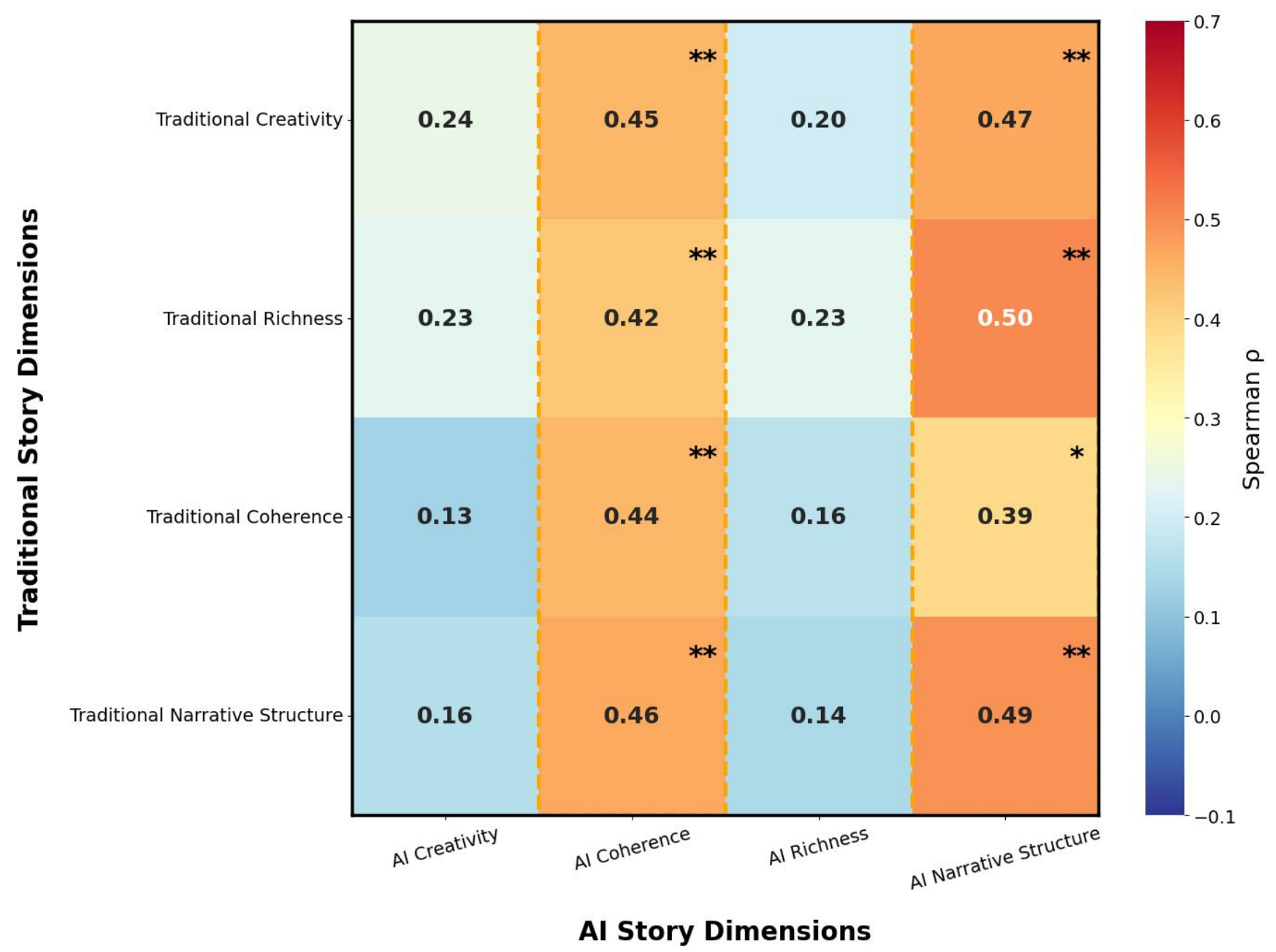
Traditional Story Dimensions
Traditional Creativity
Traditional Richness
Traditional Coherence
Traditional Narrative Structure
0.24
0.45
0.20
0.47
0.23
0.42
0.23
0.50
0.13
0.44
0.16
0.39
0.16
0.46
0.14
0.49
AI Creativity
AI Coherence
AI Richness
AI Narrative Structure
AI Story Dimensions
Spearman ρ

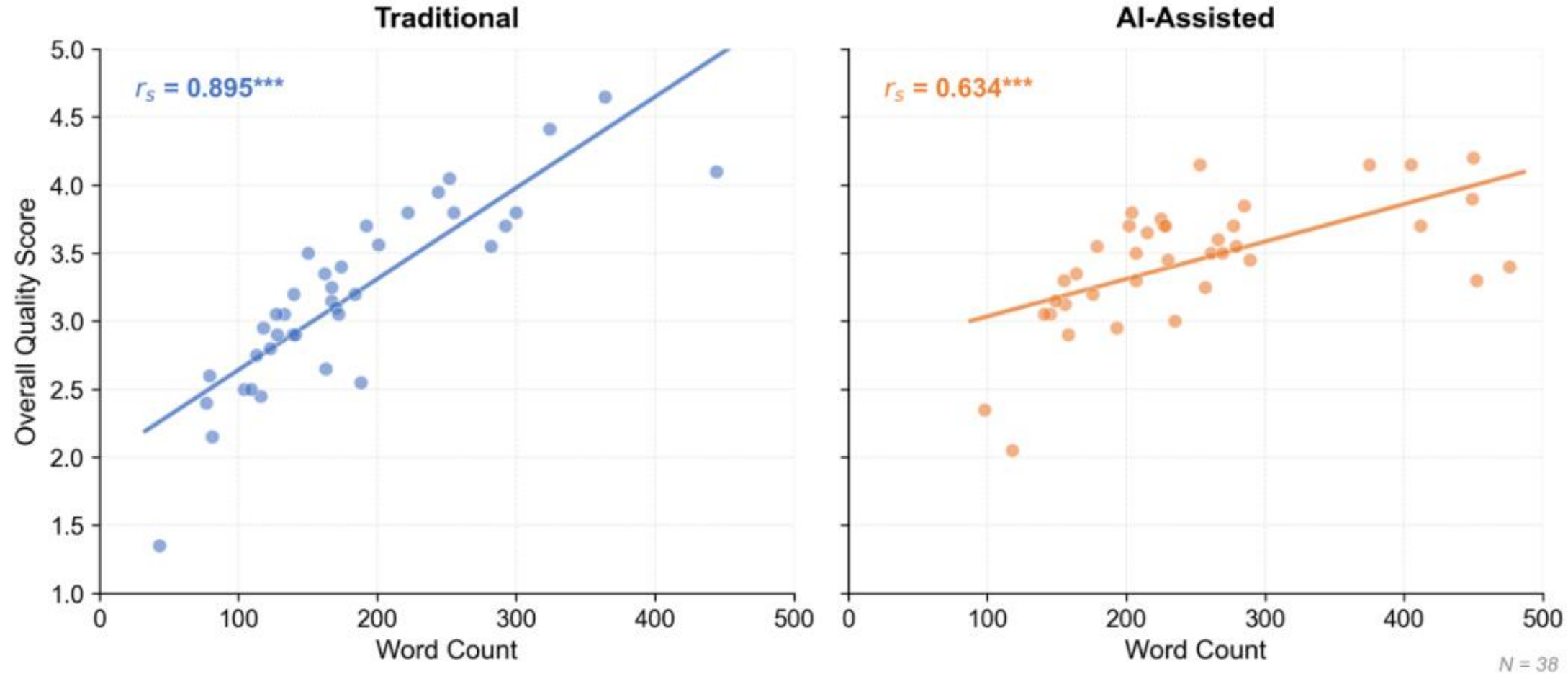
Traditional
AI-Assisted
$r_s$ = 0.895***
$r_s$ = 0.634***
Overall Quality Score
Word Count
N = 38

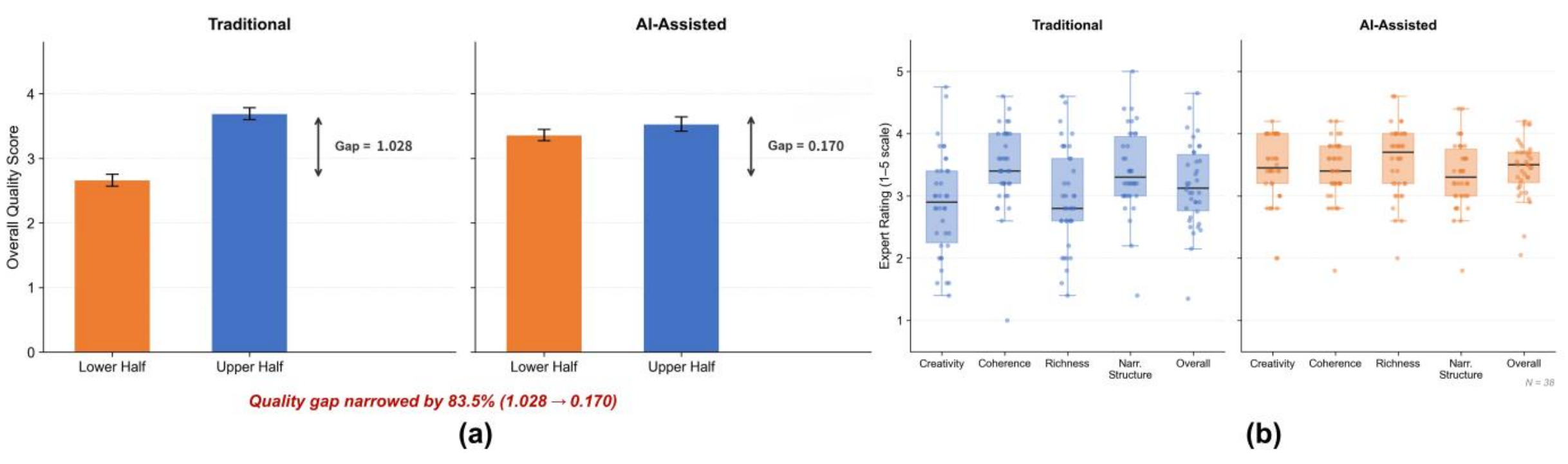
Traditional
AI-Assisted
Overall Quality Score
Gap = 1.028
Gap = 0.170
Lower Half
Upper Half
Quality gap narrowed by 83.5% (1.028 → 0.170)
Expert Rating (1–5 scale)
Creativity
Coherence
Richness
Narr. Structure
Overall
N = 38

(a)

(b)

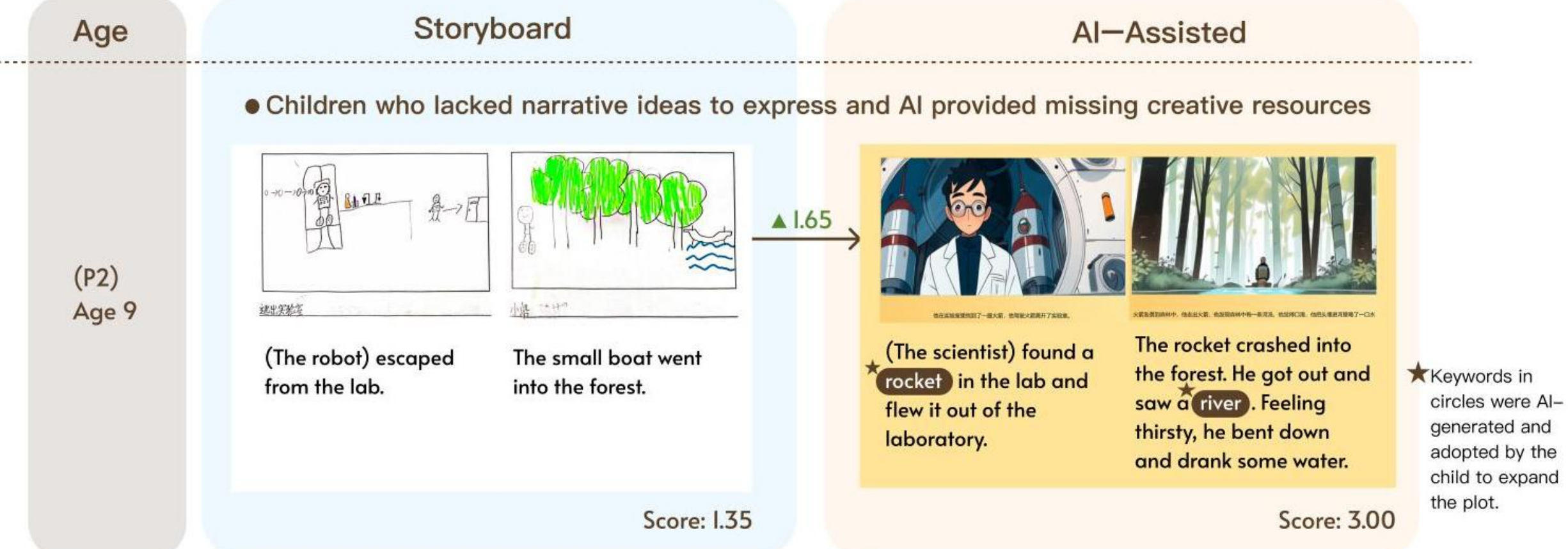
Age
Storyboard
AI–Assisted
Children who lacked narrative ideas to express and AI provided missing creative resources
(P2)
Age 9
(The robot) escaped from the lab.
The small boat went into the forest.
▲1.65
(The scientist) found a rocket in the lab and flew it out of the laboratory.
The rocket crashed into the forest. He got out and saw a river. Feeling thirsty, he bent down and drank some water.
Score: 1.35
Score: 3.00
★Keywords in circles were AI–generated and adopted by the child to expand the plot.

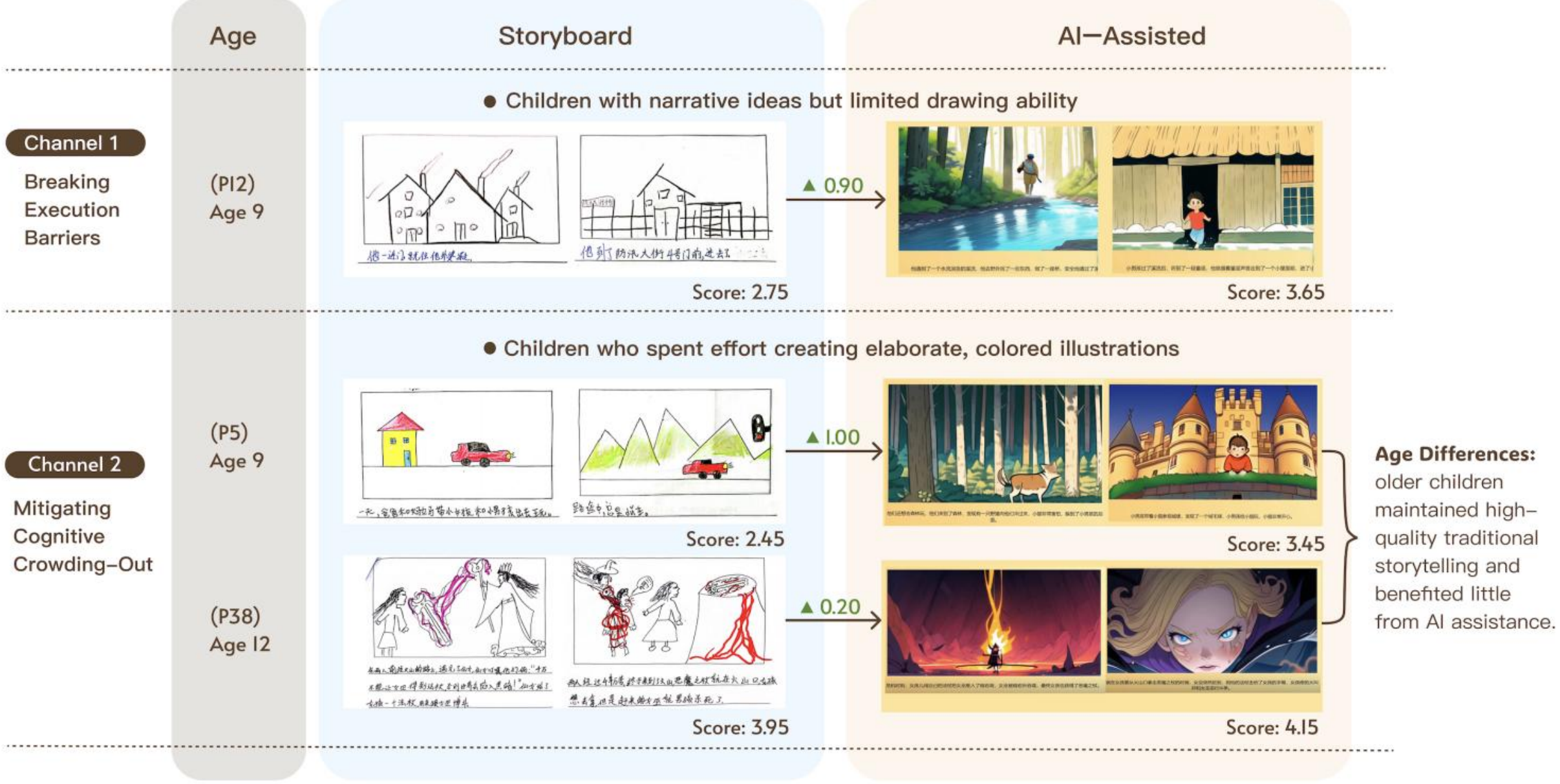
Age
Storyboard
AI–Assisted
Children with narrative ideas but limited drawing ability
Channel 1
Breaking Execution Barriers
(P12)
Age 9
▲0.90
Score: 2.75
Score: 3.65
Children who spent effort creating elaborate, colored illustrations
Channel 2
Mitigating Cognitive Crowding–Out
(P5)
Age 9
▲1.00
Score: 2.45
Score: 3.45
(P38)
Age 12
▲0.20
Score: 3.95
Score: 4.15
Age Differences: older children maintained high–quality traditional storytelling and benefited little from AI assistance.

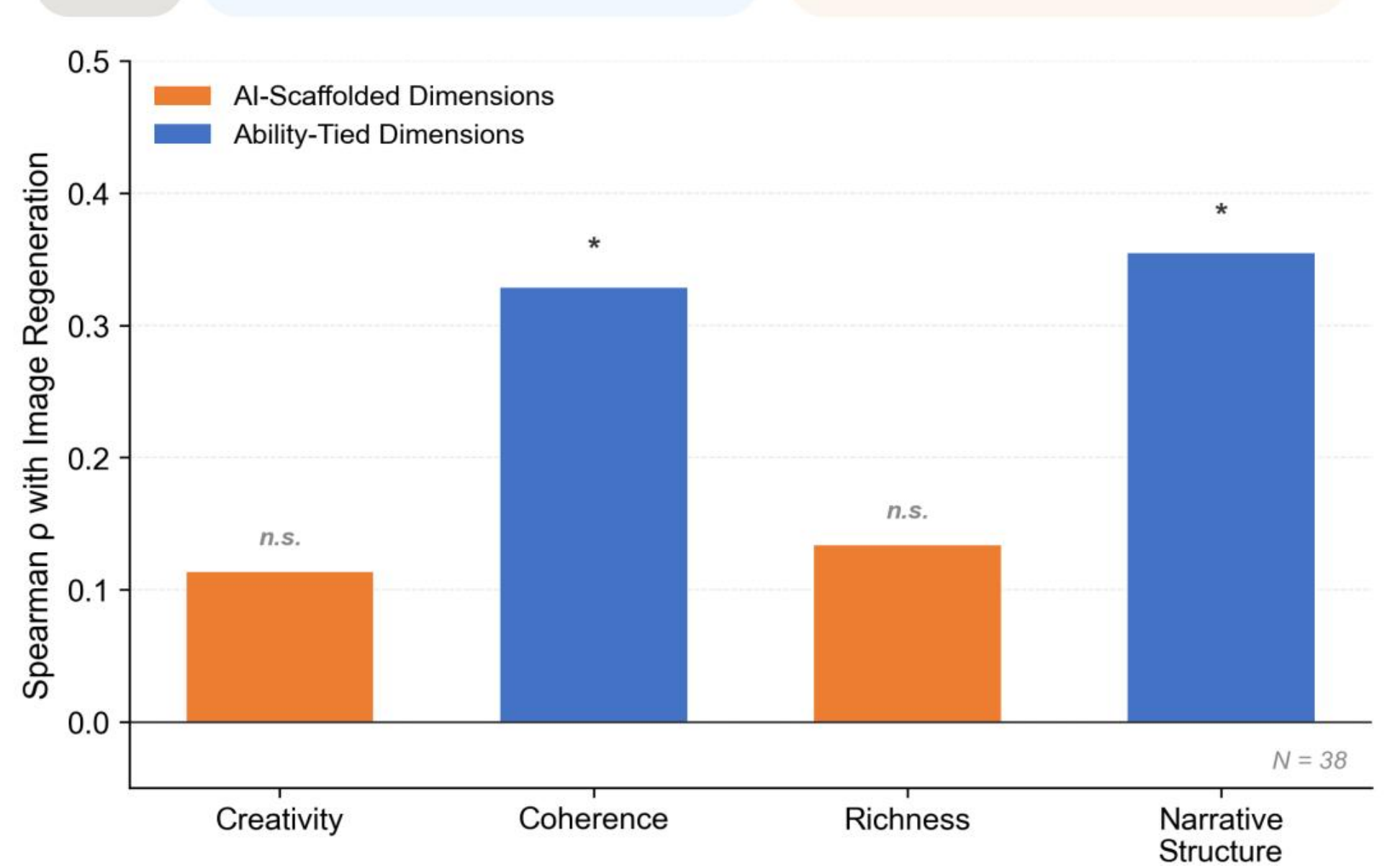
Spearman ρ with Image Regeneration
0.5
0.4
0.3
0.2
0.1
0.0
AI-Scaffolded Dimensions
Ability-Tied Dimensions
n.s.
*
n.s.
*
N = 38
Creativity
Coherence
Richness
Narrative Structure

# Figure List

**Fig. 1** Overview of the *StoryPrompt* system. (a) Phase 1 (Story Planning): children select a character, define personality traits, choose three sequential scenes, and assign emotional arcs for each scene. (b) Phase 2 (Story Creation): children compose six story paragraphs using AI-generated keywords and generate comic-style illustrations after each paragraph.

**Fig. 2** Overview of the study procedure: a counterbalanced within-subjects design conducted over two days. Each child completed a pre-study questionnaire and a storytelling instruction session, then participated in both the AI-assisted and traditional storyboard conditions (counterbalanced in order), followed by post-session interviews.

**Fig. 3** The three-stage experimental process. (a) A teacher-led storytelling instruction session for all participants. (b) The AI-assisted condition where a child uses the AI system individually. (c) The traditional storyboard condition where stories are sketched and written on paper.

**Fig. 4** Quality gain (AI minus traditional) by performance tercile and evaluation dimension. Lower-performing children (orange) show substantial gains across all dimensions, particularly in creativity and richness. Higher-performing children (blue) remain largely unchanged in creativity and richness but show notable decreases in coherence and narrative structure. Error bars indicate standard errors.

**Fig. 5** Cross-dimensional Spearman correlation matrix between traditional-condition and AI-condition story-quality dimensions ($N = 38$). Traditional-condition dimensions were significantly associated with AI coherence and AI narrative structure, whereas none was significantly associated with AI creativity or AI richness. $^{*}p < .05$, $^{**}p < .01$.

**Fig. 6** Association between text length, measured as Chinese character count, and overall story quality in the traditional and AI-assisted conditions ($N = 38$). The strong association in the traditional condition weakened under AI assistance, suggesting that story quality was less tightly tied to output volume. $^{***}p < .001$.

**Fig. 7** Floor-raising convergence pattern in story quality. (a) The median-split quality gap narrowed from 1.028 points in the traditional condition to 0.170 points under AI assistance, a descriptive reduction of 83.5%. (b) AI-condition scores also showed descriptively smaller dispersion in overall quality and creativity.

**Fig. 8** Compensatory scaffolding mechanism for a low-performing child showing narrative quality improvement (+1.65) through the adoption of AI-generated keywords (circled) as creative anchors.

**Fig. 9** Restorative release mechanism via Channel 1 (execution barrier removal, illustrated by P12) and Channel 2 (cognitive resource reallocation, illustrated by P5). P38 serves as a developmental contrast case: older children with sufficient cognitive capacity to sustain both drawing and narration simultaneously maintained high traditional quality and gained minimally from AI assistance.

**Fig. 10** Zero-order Spearman correlations between image regeneration count and four AI storytelling quality dimensions ($N = 38$). Image regeneration was significantly associated with the structural dimensions, coherence and narrative structure, but not with the content dimensions, creativity and richness. Baseline-controlled partial correlations were attenuated to non-significance, so these associations should be interpreted as exploratory. $^{*}p < .05$.